\PassOptionsToPackage{table}{xcolor}
\documentclass[sigplan, nonacm, 9pt]{acmart}
\AtBeginDocument{%
	}

\usepackage{mathrsfs, amsmath,amsfonts}
\usepackage{graphicx}
\usepackage{textcomp}
\usepackage{mdframed}

\usepackage{lipsum}

\usepackage[utf8]{inputenc}
\usepackage[T1]{fontenc}

\usepackage[abbreviations]{foreign}
\usepackage[table]{xcolor}
\usepackage{algorithm}
\usepackage{algpseudocode}
\usepackage[most]{tcolorbox}

\usepackage{soul}    

\definecolor{hostcolor}{RGB}{220, 230, 255}  
\definecolor{pimcolor}{RGB}{0, 138, 14}   
\definecolor{transfercolor}{RGB}{255, 230, 230} 

\definecolor{codegreen}{RGB}{28,172,0} 
\definecolor{codeblue}{RGB}{51,102,255} 
\definecolor{codered}{RGB}{255,51,51} 

\newtcolorbox{greenline}[1][]{%
	colback=codegreen!10, 
	colframe=codegreen!50, 
	arc=2pt, 
	boxrule=0.5pt, 
	left=0pt, right=2pt, top=2pt, bottom=2pt, 
	#1 
}

\newtcolorbox{blueline}[1][]{%
	colback=codeblue!10, 
	colframe=codeblue!50, 
	arc=2pt, 
	boxrule=0.5pt, 
	left=0pt, right=2pt, top=2pt, bottom=2pt, 
	#1 
}

\usepackage[labelfont=bf,labelsep=period]{caption}
\usepackage[nodayofweek]{datetime}
\usepackage[inline]{enumitem}
\usepackage{float}
\usepackage{geometry}
\usepackage{graphicx}
\usepackage[utf8]{inputenc}
\usepackage{times}
\usepackage{url}

\usepackage{xspace}
\usepackage{pifont}

\floatstyle{ruled}
\newfloat{algo}{htbp}{algo}
\floatname{algo}{Algorithm}

\def \ifempty#1{\def\temp{#1} \ifx\temp\empty }

\usepackage{textcomp}
\usepackage{listings}
\usepackage{booktabs} 
\usepackage{multirow}
\usepackage{tabularx}
\usepackage{makecell}
\usepackage{float}
\usepackage[export]{adjustbox}
\usepackage{eso-pic}

\usepackage{calc}

\usepackage{needspace}

\newcommand{\dpxor}{$\mathsf{dpXOR}$\xspace}

\renewcommand{\arraystretch}{1}

\newcolumntype{R}{>{\raggedleft\arraybackslash}X}
\newcolumntype{P}[1]{>{\raggedleft\arraybackslash}p{#1}}

\definecolor{prioritycolor}{HTML}{969bce}
\definecolor{darkgray}{HTML}{262626}

\usepackage{fontawesome5}
\newboolean{showcomments}
\setboolean{showcomments}{true}
\ifthenelse{\boolean{showcomments}}
{ \newcommand{\mynote}[3]{
		\fbox{\bfseries\sffamily\scriptsize#1}
		{\small$\blacktriangleright$\textsf{\emph{\color{#3}{#2}}}$\blacktriangleleft$}}}
{ \newcommand{\mynote}[3]{}}

\newcounter{numobserv} 
\definecolor{beaublue}{rgb}{0.88, 0.93, 0.93}

\usepackage[most]{tcolorbox}
\colorlet{shadecolor}{beaublue}
\tcbset{
	colback=beaublue,
	boxrule=0.5pt,
	boxsep=0pt,
	left=4pt,
	right=4pt,
	top=4pt,
	bottom=4pt,
	sharp corners,
	colframe=gray!70,
}
\newcommand{\observ}[1]{
	\addtocounter{numobserv}{1}
	\begin{tcolorbox}	
		\textit{\textbf{Take-away\,\thenumobserv\,:} #1 }	
	\end{tcolorbox}
}

\newtcolorbox{variantbox}[1]{%
	enhanced, 
	attach boxed title to top center={yshift=-2.8mm,yshifttext=-1mm,xshift=-10mm},
	colback=gray!10, 
	colframe=black, 
	colbacktitle=black, 
	sharp corners,
	top=10pt,
	fonttitle=\color{black},
	title=\textbf{#1},   
	fontupper= \ttfamily,
	boxed title style={size=small, colframe=black, colback=gray!10, sharp corners, boxrule=0.5pt} 
}

\makeatletter
\@ifclassloaded{acmart}{}{%
  \usepackage{titlesec}
  \renewcommand{\thesubsubsection}{\arabic{subsubsection}.}
  \titleformat{\subsubsection}[runin]
  {\normalfont\bfseries\itshape}
  {\thesubsubsection}{0.5em}{}[\hspace{0.5em}\\~\\]
}
\makeatother

\newcommand{\subpoint}[1]{\smallskip\noindent\textbf{#1}\xspace}

\newcommand{\subsubpoint}[1]{\smallskip\noindent\textit{#1}\xspace}

\usepackage{hyperref}
\usepackage{xspace}
\newcommand{\copyrighttext}{ \scriptsize \textcopyright 2025 ACM.               
	Personal use of this material is permitted.                                 
	Permission from ACM must be obtained for all other uses,                   
	in any current or future media, including reprinting/republishing this      
	material for advertising or promotional purposes, creating new collective   
	works, for resale or redistribution to servers or                           
	lists, or reuse of any copyrighted component of this work in other works.   
	This is the author’s version of the work. The final version is published in the proceedings of the 26th International Middleware Conference.}

\newcommand{\sys}{\textsc{IM-PIR}\xspace}

\usepackage[font=footnotesize]{caption}
\usepackage{color,soul}

\begin{document}
	
	\title{\sys: In-Memory Private Information Retrieval}

	
	
	\author{Mpoki Mwaisela}
	\affiliation{%
		\institution{University of Neuchâtel}
		\city{Neuchâtel}
		\country{Switzerland}
	}
	\email{mpoki.mwaisela@unine.ch}

	\author{Peterson Yuhala}
	\affiliation{%
		\institution{University of Neuchâtel}
		\city{Neuchâtel}
		\country{Switzerland}
	}
	\email{peterson.yuhala@unine.ch}

	\author{Pascal Felber}
	\affiliation{%
		\institution{University of Neuchâtel}
		\city{Neuchâtel}
		\country{Switzerland}
	}
		\email{pascal.felber@unine.ch}

	\author{Valerio Schiavoni}
	\affiliation{%
		\institution{University of Neuchâtel}
		\city{Neuchâtel}
		\country{Switzerland}	
	}
	\email{valerio.schiavoni@unine.ch}

	\date{\today}
	
	\begin{abstract}
Private information retrieval (PIR) is a cryptographic primitive that allows a client to securely query one or multiple servers without revealing their specific interests.
In spite of their strong security guarantees, current PIR constructions are computationally costly.
Specifically, most PIR implementations are memory-bound due to the need to scan extensive databases (in the order of GB), making them inherently constrained by the limited memory bandwidth in traditional processor-centric computing architectures.
Processing-in-memory (PIM) is an emerging computing paradigm that augments memory with compute capabilities, addressing the memory bandwidth bottleneck while simultaneously providing extensive parallelism.
Recent research has demonstrated PIM's potential to significantly improve performance across a range of data-intensive workloads, including graph processing, genome analysis, and machine learning.  
      
In this work, we propose the first PIM-based architecture for multi-server PIR.
We discuss the algorithmic foundations of the latter and show how its operations align with the core strengths of PIM architectures: extensive parallelism and high memory bandwidth.
Based on this observation, we design and implement \sys, a PIM-based multi-server PIR approach on top of UPMEM PIM, the first openly commercialized PIM architecture.
Our evaluation demonstrates that a PIM-based multi-server PIR implementation significantly improves query throughput by more than $3.7\times$ when compared to a standard CPU-based PIR approach.
	
\end{abstract}


	\newcommand{\copyrightnotice}{\begin{tikzpicture}[remember picture,overlay]       
	\node[anchor=south,yshift=2pt,fill=yellow!20] at (current page.south) {\fbox{\parbox{\dimexpr\textwidth-\fboxsep-\fboxrule\relax}{\copyrighttext}}};
	\end{tikzpicture}
	}

	\maketitle
\section{Introduction}
\label{sec:intro}
Private information retrieval (PIR)~\cite{chor95,single_pir} is a cryptographic primitive that enables a client to retrieve data from a public database without revealing the query content nor the accessed data.
It can either be \emph{single-server} PIR~\cite{cachin1999computationally, gentry2005,simplePIR, single_pir}, where cryptographic techniques like fully homomorphic encryption (FHE)~\cite{gentry09,bfv-1,bfv-2,bgv2012} are leveraged to achieve query privacy, or \emph{multi-server} PIR~\cite{DPF}, where the database is replicated across multiple non-colluding servers and techniques such as secret sharing~\cite{fss} are used to retrieve data obliviously.
PIR has been explored as a critical building block for a wide range of privacy-preserving applications including medical record retrieval \cite{pir_medical_record}, certificate transparency \cite{CA,simplePIR}, private blocklist lookup \cite{Private_web_search}, private media streaming \cite{popcorn} and verification of compromised credentials \cite{GPU-CIP}, among others.


In spite of their strong security guarantees, PIR schemes incur significant computational overhead: to preserve query privacy, the server typically processes the entirety of the database for each query~\cite{simplePIR, XPIR, sealPIR}, making PIR a fundamentally memory-bound primitive~\cite{inspire}.
As such, scaling PIR to support large databases efficiently remains challenging.
While several research efforts~\cite{simplePIR, GPU-CIP, GPU-DPF, XPIR,sealPIR} have been made to address the computational costs associated with PIR, the proposed methods are based on processor-centric compute architectures (\ie CPUs and GPUs) which remain constrained by the \emph{memory wall} -- the bottleneck stemming from the mismatch between CPU processing speed and memory bandwidth~\cite{memory-wall-mckee}.

A promising approach to mitigating the memory wall in modern computing architectures is \textit{processing-in-memory} (PIM)~\cite{elliott92, gokhale95, patterson97, draper2002}, a computing paradigm that aims to integrate computational logic, ranging from simple logic units \cite{simplePIMANDandOR} to general-purpose cores \cite{UPMEM_PIM}, directly into memory chips, enabling data to be processed in memory.
A wide range of PIM architectures have been proposed in the literature~\cite{mutlu18,hashemi16, hashemi-micro16, ahn-isca15, zhu2013accelerating, zhang-top-pim}, but only recently has the first openly commercialized PIM architecture emerged: UPMEM PIM~\cite{UPMEM_PIM}.
The latter associates each 64\,MB chunk of DRAM with a low-power core called a \emph{DRAM processing unit} (DPU), yielding an aggregate memory bandwidth of up to 2\,TB/s~\cite{gomez22}.
In addition, the large number of DPUs deployed across the memory system enables extensive parallelism.
Recent studies have exploited UPMEM PIM for data-intensive applications including graph processing~\cite{sparsep, gomez22}, genome analysis~\cite{diab2023framework, alser2020accelerating}, machine learning~\cite{gilbert24} showing large performance improvements.
 



In this work, we focus on designing a PIR scheme on top of a PIM architecture, addressing the memory bandwidth problem inherent in current processor-centric PIR schemes.
We specifically target multi-server PIR instead of the single-server variant based on FHE, as FHE involves complex algorithms like the number theoretic transform (NTT)~\cite{NTT1,NTT2,NTT3} which are poorly suited to current PIM architectures like UPMEM PIM. 
This was shown by previous studies~\cite{gilbert24, PIM_HE} that explored UPMEM PIM for FHE acceleration.
In contrast, multi-server PIR schemes are typically built upon linear secret-sharing techniques~\cite{DPF} which involve relatively lightweight operations such as bitwise XORs and dot products, which naturally map to the strengths of PIM architectures: extensive parallelism and high memory bandwidth.

We propose \sys,\footnote{Pronounced "impire".} a PIM-based multi-server PIR design which leverages the strengths of PIM architectures by offloading the core PIR server-side computations: XOR-based computations and inner products, directly onto UPMEM PIM DPUs, thereby enabling in-place query processing over large databases.
This reduces data movement between memory and compute units, and enhances performance for multi-server PIR schemes.
By aligning algorithmic structure with architectural strengths, our PIM-based multi-server PIR demonstrates the potential of emerging memory-centric platforms like UPMEM's for practical privacy-preserving processing operations.
To the best of our knowledge, our work is the first to propose a thorough PIM-based design for multi-server PIR.

Overall, our paper provides the following contributions:
\begin{itemize}
\item We design \sys, a high-throughput PIM-based approach for accelerating multi-server PIR query processing.
\item We implement and evaluate \sys on UPMEM PIM, the first commercialized architecture. Our experimental results show \sys improves query throughput and latency by more than $3.7\times$ compared to state-of-the-art processor-centric multi-server PIR designs.
\end{itemize}

%
%

\subpoint{Roadmap.}  
The remainder of this paper is structured as follows: 
\S\ref{sec:background} introduces key concepts, including private information retrieval protocols and the UPMEM processing-in-memory architecture.  
\S\ref{sec:arch} discusses our PIM-based multi-server PIR design, while \S\ref{sec:implem} outlines the corresponding implementation details.
We evaluate \sys in \S\ref{sec:evaluation}, providing a performance comparison of our solution againsts current processor-centric PIR designs.
We then discuss related work in \S\ref{sec:rw} and conclude our paper in \S\ref{sec:conclusion}.

\section{Background}
\label{sec:background}

In this section, we provide a thorough background on private information retrieval techniques and discuss the issues faced with traditional processor-centric PIR implementations (\S\ref{sec:pir}).
We then provide an overview of processing-in-memory, and UPMEM's PIM architecture (\S\ref{sec:pim}).

\subsection{Private Information Retrieval}
\label{sec:pir}

\emph{Private information retrieval} (PIR)~\cite{chor95,single_pir} is a fundamental privacy-preserving primitive that allows a client to retrieve a specific item from a \emph{public database} without the database server learning any information about the client's query nor the item being accessed. 
In a typical PIR setting, we consider a database \( D \) consisting of \( N \) items; the goal is to obtain the \( i \)-th element \( D[i] \) while keeping the index \( i \) confidential. 
This privacy guarantee is of paramount importance in a variety of applications that require the privacy of user queries~\cite{CA,pir_medical_record,popcorn,Private_web_search,PrivateDB}. 

PIR techniques are broadly classified into two categories: \emph{single-server} PIR and \emph{multi-server} PIR. 
In single-server PIR, query privacy is achieved by relying on cryptographic hardness assumptions, \ie ring learning-with-errors problem (RLWE)~\cite{rlwe} and techniques like fully homomorphic encryption (FHE)~\cite{gentry09,bfv-1,bfv-2,bgv2012} which build upon these assumptions.
On the other hand, in multi-server PIR the database is replicated across multiple non-colluding servers.
The client then sends different queries to each of the servers and derives the requested item by (mathematically) combining the responses from the servers.
The privacy of the user's query is achieved through the non-collusion assumption, rather than cryptographic hardness.


A central aspect of PIR protocols is the so-called \emph{all-for-one} principle~\cite{chor95,PIR_computations,XPIR,sealPIR}, which requires that the server (or servers) process the entire database for every query. 
This exhaustive processing is not an incidental detail but a deliberate design choice to ensure that the data access pattern remains statistically independent of the specific query being made. 
By requiring a full sweep of the database, the PIR protocols prevent any leakage of information that could otherwise be exploited to infer the user's interest by the untrusted server. 
While the extensive data processing inherent in the \emph{all-for-one} approach is critical for achieving strong privacy, it simultaneously poses significant challenges in terms of computational load and memory bandwidth, issues that are particularly pronounced in large-scale deployments. Consequently, a considerable body of research is devoted to optimizing this trade-off, striving to enhance system efficiency while preserving the robust privacy characteristics that define PIR systems~\cite{CA,fss,GPU-CIP,GPU-DPF,inspire,popcorn,sealPIR,XPIR,simplePIR}.

\subpoint{Basic notation.}
In the remainder of this work, we adopt the following notations.
$D$ represents a public PIR database with $N$ items.
When used, $L$ is the vector length (\eg number of bits) of each database entry, and $\mathbb{F}_p$ is an integer field with modulus $p$.
Vectors are formatted in bold, \eg $\mathbf{v}$ and the \emph{i}-th index is denoted by $\mathbf{v}[i]$.
$\oplus$ represents the exclusive-OR, \ie XOR bitwise operation.\footnote{XOR represents the addition operation in a binary field.}
Henceforth, we use the terms database and DB interchangeably.
We use the symbols KB, MB, and GB to denote $2^{10}$, $2^{20}$, and $2^{30}$ bytes respectively.

\subsection{Single-Server PIR}
\begin{figure}[!t]
	\centering
	\includegraphics[width=1\linewidth]{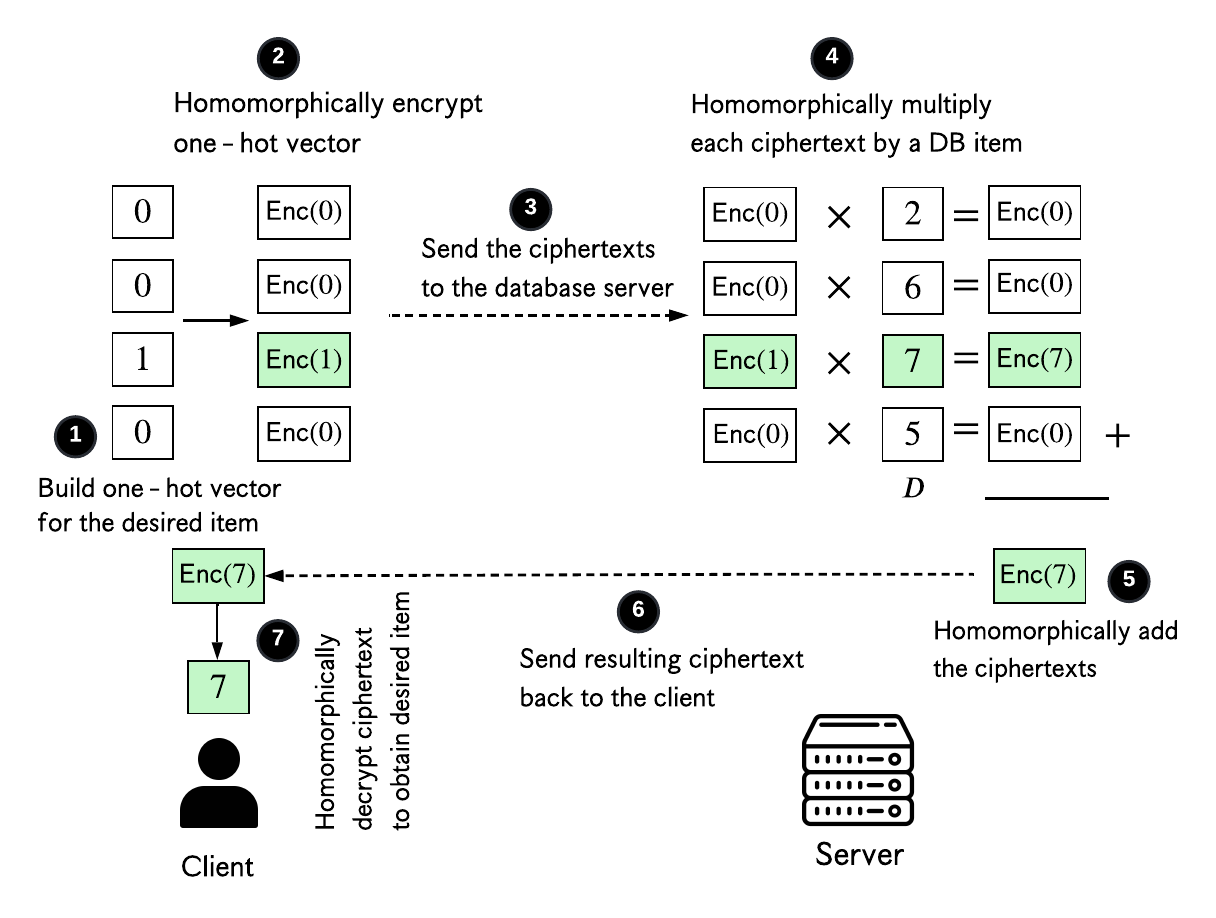}
	\caption{Single-server PIR with fully homomorphic encryption. To fetch  record 7 from a 4-entry database $D=[2,6,7,5]$, the client homomorphically encrypts a one-hot query vector into a query ciphertext which is sent to the database server. The server first homomorphically multiplies the query ciphertext with the entire DB, and then homomorphically adds the resulting ciphertexts to obtain $\mathsf{Enc}(7)$. This is sent back to the client who decrypts it to obtain the record of interest, 7, obliviously.}
	\label{fig:ss-pir}
\end{figure}
\autoref{fig:ss-pir} presents an overview of a simple FHE-based single-server PIR.
Given a public database $D=[2,6,7,5]$ containing four items, a client wishes to retrieve the $3^{rd}$ item from D, \ie $D[2]=7$ in a privacy-preserving way.
The client generates a query $q = [0, 0, 1, 0]$~\ding{202} which is a one-hot vector (\ie indicator vector) of length 4, where the $3^{rd}$ slot in the vector is 1 and the rest 0.
To preserve query privacy, the client encrypts ($\mathsf{Enc}$) each slot of the vector using FHE~\ding{203} to produce a vector of ciphertexts $[c_0, c_1, c_2, c_3] = [\mathsf{Enc}(0), \mathsf{Enc}(0), \mathsf{Enc}(1), \mathsf{Enc}(0)]$.
The client then sends this query ciphertext to the server~\ding{204}, which cannot learn anything about the actual item being requested.
The server homomorphically multiplies each $c_i$ by the corresponding DB item $D[i]$ to obtain a vector of ciphertexts~\ding{205}, each encrypting either 0 or the desired DB item.
That is, the server does:
\[
[c_0, c_1, c_2, c_3]\circ[2,6,7,5] = [\mathsf{Enc}(0), \mathsf{Enc}(0), \mathsf{Enc}(7), \mathsf{Enc}(0)]
\]
where $\circ$ represents a point-wise FHE multiplication of both vectors.\footnote{In fully homomorphic encryption, if a ciphertext $c_1$ encrypts a plaintext $p_1$ and we have another plaintext $p_2$, then $c'= c_1*p_2$ encrypts the value $p_1*p_2$}
The server then homomorphically adds the ciphertexts~\ding{206}, resulting in a single ciphertext encrypting the desired DB item:
$\mathsf{Enc}(0) + \mathsf{Enc}(0) + \mathsf{Enc}(7) + \mathsf{Enc}(0) = \mathsf{Enc}(7)$, where $+$ represents FHE addition.
This single ciphertext result, $\mathsf{Enc}(7)$ is sent to the client~\ding{207} which decrypts ($\mathsf{Dec}$) the ciphertext~\ding{208} to obtain the desired plaintext item, $D[2] = \mathsf{Dec}(\mathsf{Enc}(7)) = 7$.

\subpoint{Computational complexity.}
Following from the illustrative example above, the main computations in single-server PIR are FHE multiplication and addition operations.
FHE addition is simple, and results in a time complexity $\mathcal{O}(N)$ for vectors of size $N$.
In contrast, FHE multiplication is more costly, requiring expensive convolution operations.
The number theoretic transform (NTT)~\cite{NTT2} is commonly used to achieve such homomorphic multiplications with a time complexity of $\mathcal{O}(NlogN)$ for vectors of length $N$.
Since the PIR protocol performs computations on the entire database for each query (\ie all-for-one principle), the resulting computational overhead is very large.

\subsection{Multi-Server PIR}
\begin{figure} [!t]
	\centering
	\includegraphics[width=1\linewidth]{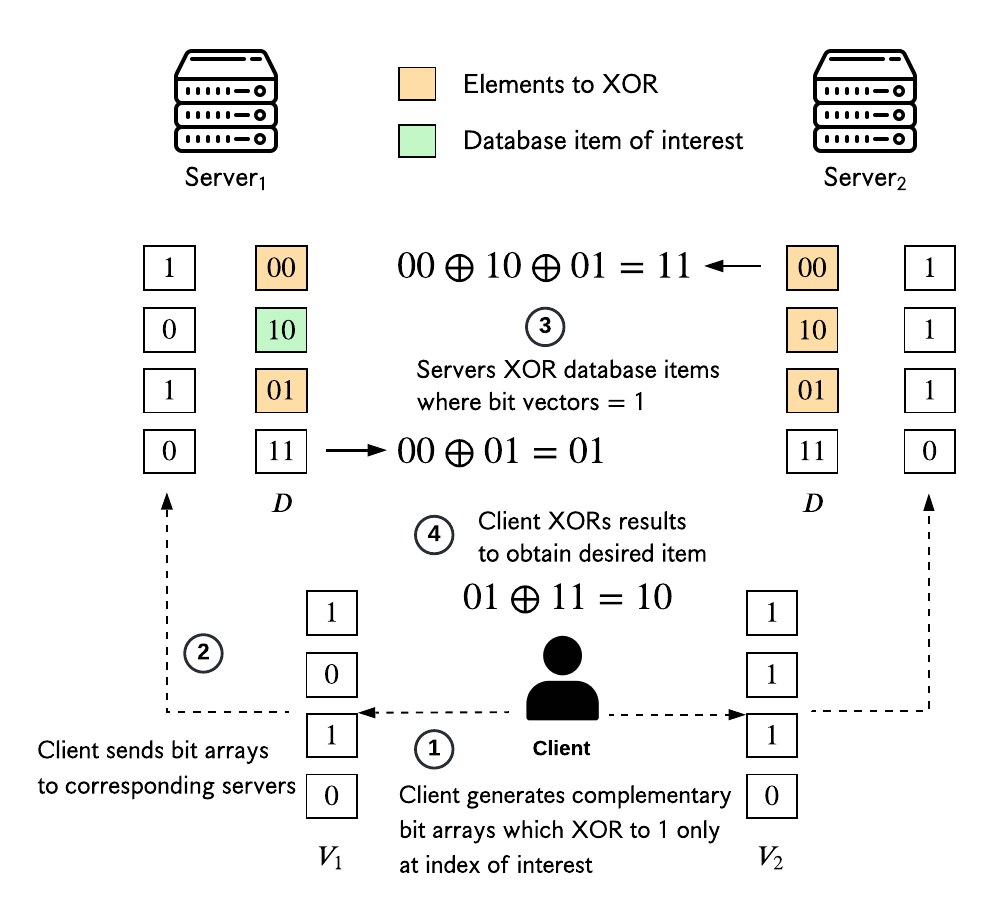}
	\caption{Multi-server PIR with $n=2$ non-colluding servers. To fetch record $10$ from the replicated database $[00, 10, 01, 11]$, the client generates (random) bit vectors which XOR to 1 only at index of interest (\ie $2$ in this case). The servers XOR elements of the database where the bit vectors have value 1. The results are sent to the client which obtains the desired item by XORing both results: $01\oplus 11 = 10$. }
	\label{fig:ms-pir}
\end{figure}


In a multi-server PIR setting, query privacy is not achieved with FHE, but by secret-sharing the query across $n$ replicated servers ($n \geq 2$). 
Privacy is guaranteed under the assumption that these servers do not collude. 
This non-collusion assumption is standard in multi-server PIR and is often credible in practice when servers are operated by independent entities with contractual separation, auditing, and distinct key management \mbox{\cite{GPU-CIP,Prio}}.

Formally, in multi-server PIR, we consider a client C who seeks to privately access item $D[i] \in \mathbb{F}_p^L$ from a database (or table) $D\in \mathbb{F}_p^{N\times L}$ which is replicated (or duplicated) across two non-colluding servers, $S_1$ and $S_2$.
A simple (naive) approach for the client to proceed is encode their query as two random vectors $\mathbf{v_1}\in \mathbb{F}_p^N$ and $\mathbf{v_2}\in \mathbb{F}_p^N$ such that $\mathbf{v_1} \oplus \mathbf{v_2}$ is a one-hot indicator vector whose entries are all $0$, except at the index of interest $i$, where it is $1$.
Upon receiving the vectors, $S_1$ and $S_2$ individually compute a linear combination of the the database entries weighted by their query vectors ($\mathbf{v_1}$ or $\mathbf{v_2}$), yielding two subresults $r_1$ and $r_2$ which are sent back to the client. 
The client then obtains the desired time as $D[i] = r_1 + r_2$.
 
\autoref{fig:ms-pir} presents an overview of this simple multi-server PIR scheme in $\mathbb{F}_2$.
Given a public database $D$ containing four items represented as bit strings: $[00, 10, 01, 11]$ replicated across two non-colluding servers $S_1$ and $S_2$, a client wishes to retrieve the $2^{nd}$ item from D, \ie $D[1]=10$, in a privacy-preserving way.
First, the client generates two random bit vectors: $\mathbf{v_1} \in \mathbb{F}_2^4$ and $\mathbf{v_2} \in \mathbb{F}_2^4$~\ding{192} such that $\mathbf{v_1}[i] \oplus \mathbf{v_2}[i] = 0$ at all entries, except at the index/slot of interest, 2 in this case, where $\mathbf{v_1}[i] \oplus \mathbf{v_2}[i] = 1$.
For example, if $\mathbf{v_1} = [1, 0, 1, 0]$, then $\mathbf{v_2} = [1, 1, 1, 0]$. 
The client sends $\mathbf{v_1}$ to $S_1$ and $\mathbf{v_2}$ to $S_2$~\ding{193}.
Each server XORs all values in $D$ where its bit vector is 1~\ding{194}: thus, $S_1$ computes $r_1 = 00\oplus 01 = 01$ and $S_2$ computes $r_2 = 00\oplus 10\oplus 01 = 11$.
The subresults are then sent to the client, which XORs both subresults received~\ding{195} to retrieve the item of interest: $D[1] = r_1 \oplus r_2 = 01 \oplus 11 = 10$.

This idea is formalized using the notion of a \emph{distributed point function} (DPF)~\cite{DPF}, which we explain in the following.


Given 2 values $a$ and $b$, a point function $P_{a,b}(x)$ is given by:
\[
P_{a,b}(x) = \begin{cases}
b, & \text{if } x = a \\
0, & \text{otherwise}
\end{cases}
\] 
That is, P is $0$ everywhere except at $a$, where its value is $b$.

A \emph{distributed point function} (DPF) allows secret-sharing of a point function.
It represents a point function $P_{a,b}$ using two keys $k_1$ and $k_2$.
Each key individually hides $a$ and $b$, but there exists an efficient algorithm $\mathsf{Eval}$ such that: 
\[
\mathsf{Eval}(k_1, x)\oplus \mathsf{Eval}(k_2, x) = P_{a,b}(x),~~\forall x
\]
If we let $f_k$ denote the function $\mathsf{Eval}(k,\cdot)$, the functions $f_{k_0}$ and $f_{k_1}$ can be viewed as an \emph{additive secret sharing} of $P_{a,b}$~\cite{DPF}.

Based on the above discussion, we can formally define a DPF as a pair of functions $(\mathsf{Gen}, \mathsf{Eval})$, where $\mathsf{Gen}$ is a key generation function which takes values $a$ and $b$ as inputs: $\mathsf{Gen}(a,b)$, and outputs DPF keys (\ie secret shares) $k_1, k_2$ such that no individual key reveals either $a$ or $b$. 
The evaluation function, $\mathsf{Eval}$, then takes a key $k$ and an input $x$, producing a share of the function $f_{k}(x) = \mathsf{Eval}(k, x)$.


To achieve multi-server PIR (\eg for $n=2$ servers), the client can encode their query using a DPF.
Given a database $D$ containing $N$ items (bit strings), suppose a client wishes to retrieve the item at index $i^*$ = $D[i^*]$.
The client encodes their query as a point function $P_{i^*,1}$ (essentially a one-hot vector) such that: 

\[
P_{i^*,1}(j) = \begin{cases}
1, & \text{if } j = i^* \\
0, & \text{otherwise}
\end{cases}
\]
The client uses a DPF generation function to produce two keys: $\mathsf{Gen}(i^*,1) = (k_1, k_2)$, such that: 

\[
P_{i^*,1}(j) = \mathsf{Eval}(k_1, j)\oplus \mathsf{Eval}(k_2, j),~~\forall j
\]
The client sends $k_1$ to the first server and $k_2$ to the second server.
As discussed previously, $\mathsf{Eval}(k, j)$ can be seen as a function share $f_k$ of the point function.
Each server thus runs the $\mathsf{Eval}$ function with their respective keys on each database index $j\in [0,N-1]$ to obtain a vector $q = [f_k(0), f_k(1), \dots f_k(N-1)]$, and then computes a linear combination, \eg XOR, of the database entries weighted by this vector to obtain a subresult $r$:
\begin{align*}
r &= f_0\cdot D[0] \oplus \dots \oplus f_{N-1}\cdot D[N-1]\\
  &= \mathsf{Eval}(k,0)\cdot D[0] \oplus \dots \oplus \mathsf{Eval}(k,N-1)\cdot D[N-1]\\
  &= \bigoplus_{j=0}^{N-1} \mathsf{Eval}(k,j)\cdot D[j]\\
\end{align*}
Each subresult is sent to the client which XORs them to reconstruct the desired item: $D[i^*] = r_1 \oplus r_2$

Going back to the illustrative example described in \autoref{fig:ms-pir} with $D = [00, 10, 01, 11]$ and the client wishing to obtain $D[1] = 10$, the one-hot vector representing the query is $[0, 1, 0, 0]$ which corresponds to the point function $P_{1,1}$.
A DPF $=(\mathsf{Gen}, \mathsf{Eval})$ can be used to generate keys and function shares as follows: 
\begin{align*}
(k_1, k_2)\leftarrow \mathsf{Gen}(1,1)\\
P_{1,1}(x) = \mathsf{Eval}(k_1, x) \oplus \mathsf{Eval}(k_2, x), \forall x\in \{0, 1, 2, 3\}\\
\mathbf{v_1} = \left[\mathsf{Eval}(k_1, x) \right]_{x = 0}^{3} = [1, 0, 1, 0]\\
\mathbf{v_2} = \left[\mathsf{Eval}(k_2, x) \right]_{x = 0}^{3} = [1, 1, 1, 0]\\
\mathbf{v_1}\oplus \mathbf{v_2} = [0, 1, 0, 0]
\end{align*}

These shares of the query vector are then used as previously described to obtain the XOR-derived subresults which are used by the client to reconstruct the desired DB item.  

\subpoint{Computational complexity.}
The key generation function $\mathsf{Gen}$ is a relatively lightweight computation with complexity of $\mathcal{O}(log(N))$, where N is the size of the DB.
This function is performed on the client.
On the contrary, the key evaluation function $\mathsf{Eval}$ is more expensive, as it evaluates all indices of the database.
However, the most expensive operations are the dot/inner product and XORing (henceforth $\mathsf{dpXOR}$ ) of database items which (in general) require $\mathcal{O}(N)$ computation since every entry of the database is processed.
These computations represent the major bottleneck in multi-server PIR~\cite{GPU-CIP} which our work aims to tackle.

 \begin{figure} [!t]
    \centering
    \includegraphics[width=1\linewidth]{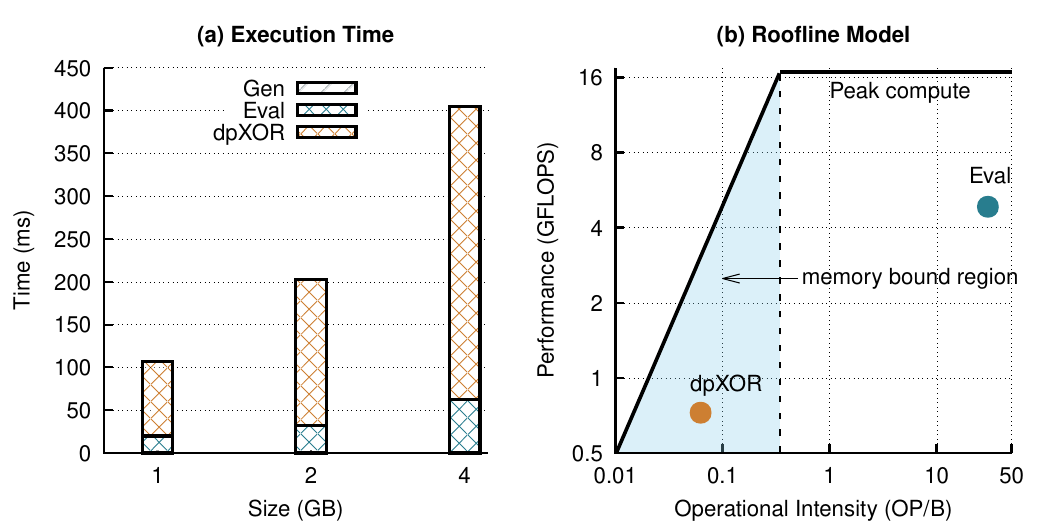}
    \caption{Breakdown of execution times for DPF-based multi-server PIR operations. Memory-bound (low operational intensity) XOR-based operations represent the primary bottleneck, motivating the need for memory-centric compute solutions.}
    \label{fig:dpf_execution}
\end{figure}
In \autoref{fig:dpf_execution}, we analyze the computational overhead of these DPF-based PIR operations across databases of varying sizes.
While client-side key generation remain relatively lightweight, the server side operations: key evaluation and $\mathsf{dpXOR}$ operations become significantly more demanding as the database size increases.
For example, a single query on a 4\,GB database reveals that $\mathsf{dpXOR}$ operations take $\approx10\times$ longer than key evaluation, which itself is $\approx1000\times$ than key generation.
Overall, processing a 4\,GB database takes about $3$s on the server, with the bulk of the time spent on $\mathsf{dpXOR}$ operations which involve accessing the entire database.
The roofline model in \autoref{fig:dpf_execution} (b) indicates that these XOR-based operations achieve low operational intensity, meaning they are memory-bound.\footnote{Memory bound workloads have low operational intensity, and compute bound workloads have high operational intensity.}

In processor-centric architectures, such operations are primarily bottlenecked by memory bandwidth limitations and the cost of moving data between the processor and memory~\cite{simplePIR}.
These performance constraints underscore the importance of addressing memory bottlenecks for practical deployment of PIR systems handling large and frequently queried databases~\cite{GoogleCT,ChromiumCT, Etherscan}.

\subsection{Processing-in-memory}
\label{sec:pim}
Processing-in-memory (PIM)~\cite{elliott92, gokhale95, patterson97, draper2002} is a memory-centric computing paradigm that integrates either general-purpose cores or specialized accelerators close to or within memory arrays. 
This design helps alleviate the data movement bottleneck caused by the costly data transfers between processors and memory in traditional processor-centric architectures, which lead to significant performance losses and energy overhead~\cite{google_workloads,PRIM}. 
Additionally, PIM addresses the growing performance gap between fast processors and slower memory modules. 
Several PIM architectures exist, including HBM-PIM~\cite{samsung_hbm_pim}, UPMEM-PIM~\cite{UPMEM_PIM,upmem}, Samsung AxDIMM~\cite{samsung_axdimm_open_innovation}, and SK Hynix AiM~\cite{sk_hynix_gddr6_pim}. 
In our work, we focus on the UPMEM PIM architecture; however, the design we propose extends to other PIM systems that adopt a similar model where the CPU executes the host applications
and PIM cores accelerate data-intensive, memory-bound operations.

\begin{figure}[!t]
    \centering    
    \includegraphics[width=1\linewidth]{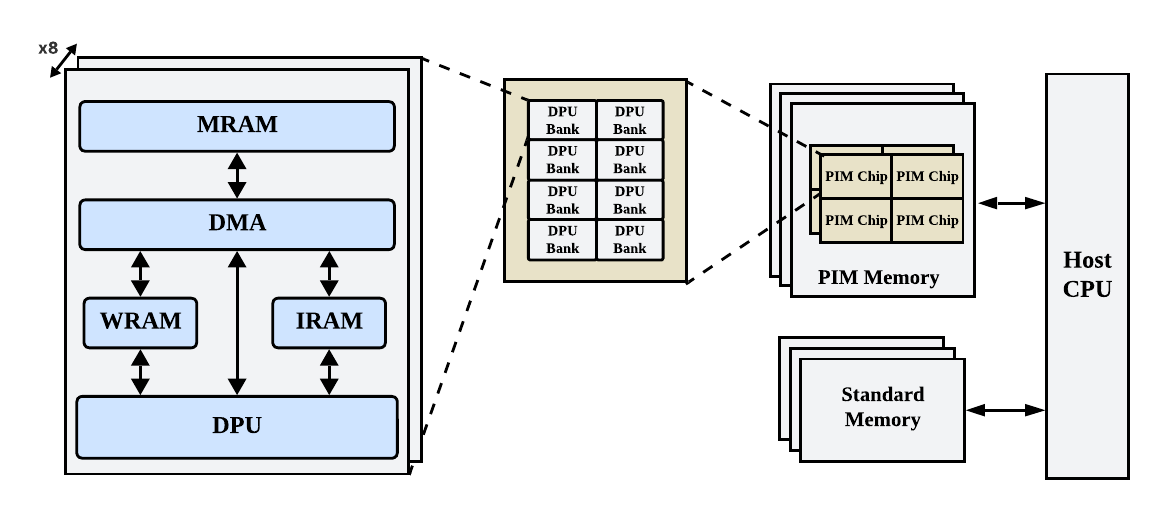}\Description{Diagram of the UPMEM PIM System hardware architecture.}
    \caption{High-level overview of the UPMEM PIM hardware architecture. The host CPU interacts with both standard main memory and PIM-enabled memory. The latter contains low-power DRAM processing units (DPUs), enabling computation in memory.}
    \label{fig:pim}
\end{figure}

\subpoint{UPMEM PIM architecture.} 
A UPMEM PIM-enabled system, as depicted in \autoref{fig:pim}, consists of a host CPU (a standard multicore processor), standard DRAM modules, and PIM-enabled memory modules. 
Each PIM module comprises two DRAM ranks, with each rank containing eight PIM chips.
Each PIM chip integrates eight DRAM Processing Units (DPUs), where each DPU is associated with 64\,MB of private main RAM (MRAM). 
In an 8\,GB memory module, this architecture results in a total of 128 DPUs. 
At the time of this writing, a UPMEM PIM-enabled system can scale to support up to 20 PIM-enabled modules, which collectively accommodate 2,560 DPUs with a total memory capacity of 160 GB.

Though a UPMEM PIM-enabled system incorporates two distinct types of memory modules (PIM-enabled or standard DRAM), it retains the standard DRAM as main memory for the host CPU. 
The host CPU offloads computations to PIM-enabled memory.
Initially, data to be processed in PIM modules is stored in standard DRAM.
It is then explicitly copied to PIM memory via the CPU.
Once PIM computations are complete, the results can be retrieved by the host CPU back into standard DRAM.

\subpoint{DPU architecture.} 
Each DPU is a general-purpose 32-bit in-order core which implements a RISC-style instruction set architecture. 
It contains $88$KB of SRAM which is split into 24KB instruction memory called instruction RAM (IRAM), and 64KB scratchpad memory called working RAM (WRAM).
Data is exchanged between WRAM and MRAM via DMA.
The DPU supports up to $24$ hardware threads to achieve scaling, with all threads sharing WRAM space.
At the time of this writing, UPMEM PIM-DPUs can operate either a 350\,MHz or 400\,MHz.
The maximum possible MRAM-WRAM bandwidth for each configurations is respectively, 700\,MB/s and 800\,MB/s, resulting in an aggregate memory bandwidth of up to 2\,TB/s (for DPUs at 350\,MHz) or 1.79\,TB/s (for DPUs at 400\,MHz)~\cite{gomez22, PIM-TREE}.

\subpoint{UPMEM PIM programming model.}
A UPMEM PIM program consists of a host program and a DPU program.
The host program executes on the CPU while the DPU program runs on the DPUs.
UPMEM provides a software development kit (SDK) allowing developers to write PIM-based software.
The program to be executed on the DPUs must be written in the C programming language, but UPMEM has announced support for Rust in the near future.
On the other hand, the host program can be written in C/C++, Python, and Java. 
The primary roles of the host program include: allocating DPUs, loading the compiled DPU binaries, and initiating DPU execution.

DPUs follow the single program multiple data (SIMD) programming model, and 24 software threads called \emph{tasklets} map to the DPU's hardware threads.
Developers are responsible for partitioning the workloads and managing synchronization across these tasklets. 

Due to architectural and design constraints related to memory access, UPMEM PIM maintains a separate address space for PIM memory and standard DRAM memory.
As a result, data to be processed by PIM DPUs must be explicity copied from DRAM to DPU MRAM.
Also, DPUs have no memory management unit (MMU), so programmers must derive the physical memory addresses in MRAM of data to be copied to the DPU~\cite{pim-mmu}.

\subpoint{This work.}
We aim to leverage PIM to tackle the data movement problem in PIR schemes, which as we have seen, are highly memory-bound.
We recall that single-server PIR computational overhead is dominated by FHE.
Recent studies~\cite{PIM_HE, gilbert24} have explored using PIM architectures like UPMEM's to accelerate FHE, but their findings indicate that current PIM architectures like UPMEM PIM are poorly adapted for FHE.
This is due to complex operations like the number theoretic transform~\cite{NTT1,NTT2,NTT3} which induce costly inter-DPU communication~\cite{PIM_HE, gilbert24}.
In contrast, multi-server PIR protocols typically rely on lightweight operations like XORs, dot products, and basic linear algebra which align well with the highly parallel and high-bandwidth design provided by current PIM architectures.
For this reason, our work focuses on accelerating multi-server PIR using PIM. 

\observ{
	Multi-server PIR algorithms are better adapted for PIM than their single-server PIR counterparts.
}

\section{Design of \sys}
\label{sec:arch}
\begin{figure*} [!t]
	\centering
	\includegraphics[scale=0.65]{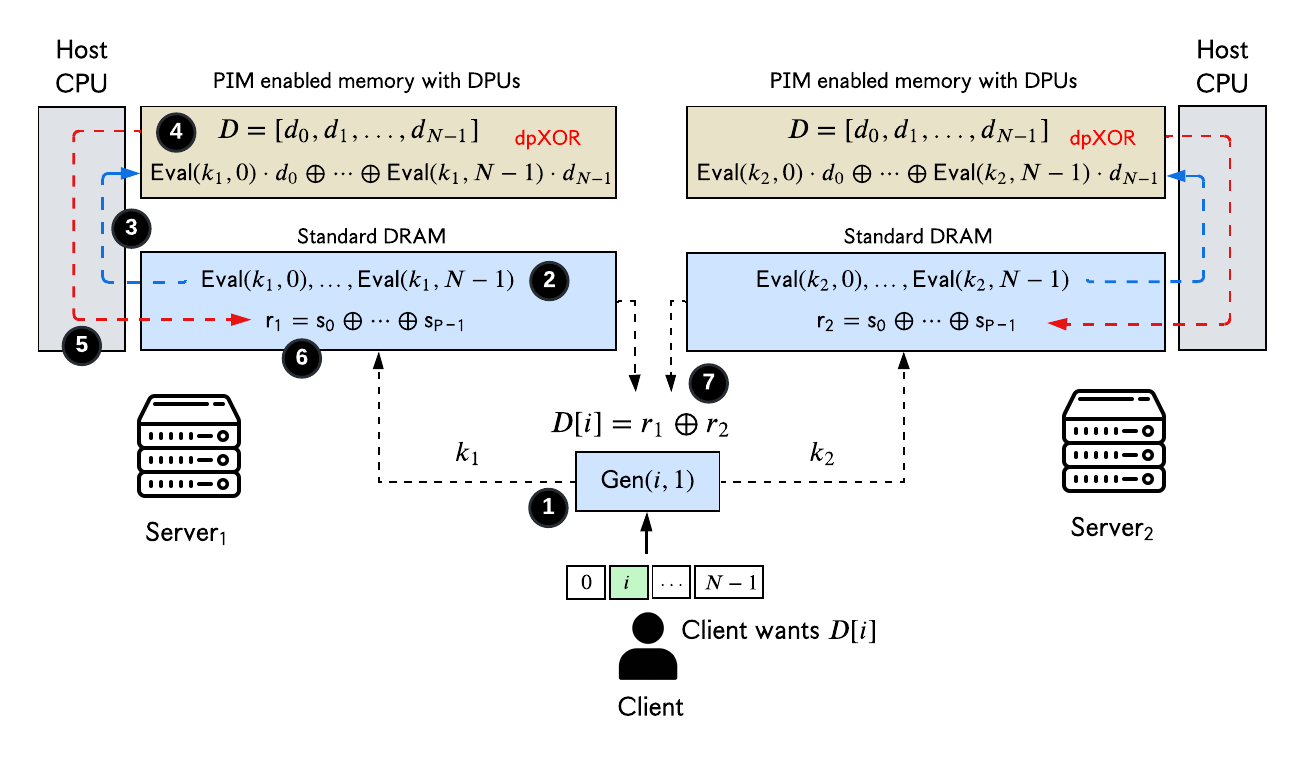}
	\caption{Multi-server PIR with \sys. Server-side computations are partitioned between the host CPU and PIM DPUs: the host performs initial DPF key evaluation, while the more costly memory-bound linear $\mathsf{dpXOR}$ operations are performed on PIM DPUs.}
	\label{fig:pir-pim}
\end{figure*}
In this section we present the design of \sys, a multi-server PIR solution that leverages a PIM architecture to overcome the limitations of current processor-centric multi-server PIR solutions.
We consider a DPF-based multi-server PIR scheme and describe how we align its algorithmic structure with the architectural strengths of both PIM and traditional CPUs.
Our design strategically partitions the core server-side PIR workload between the CPU and PIM cores: the host CPU handles initial DPF key evaluation, while the memory-bound linear operations, such as inner products and bitwise XOR operations on the entire database, are offloaded to PIM DPUs for efficient in-place parallel processing.


We consider a DPF-based multi-server PIR scheme where a client aims to retrieve an item from a public database $D = [d_0, \dots, d_N]$ replicated across two servers $S_1$ and $S_2$.
We adopt the DPF construction from \mbox{\cite{GPU-DPF}}, which extends the earlier design of \mbox{\cite{DPF}}.
These DPF constructions leverage a Goldreich-Goldwasser-Micali (GGM)~\cite{ggm-86}-based pseudorandom function (PSF) for computing DPF function shares.

While our work considers a two-server PIR construction, \ie., $n = 2$, the details are easily generalizable to multi-server PIR constructions where $n > 2$.
However, communication overhead from distributing queries increases with the number of servers.
\autoref{fig:pir-pim} illustrates the overall system design of \sys, and we detail its key steps in the rest of this section.

\subsection{DPF key generation}
Key generation is the first step in multi-server PIR and allows the client to encode their query into keys which will be used to create function secret shares used on the database servers.
In \sys, key generation function $\mathsf{Gen}$ generates the two DPF keys~\ding{202} from two inputs, the database index of interest $i$ and a security parameter $\lambda$ to produce two keys: 
\[
\mathsf{Gen}(1^\lambda, i\in 0,\dots,N-1)\rightarrow (k_1,k_2)
\]
The security parameter, $\lambda$, typically corresponds to the bit-length of a cryptographic key.
Each key consists of two 2-dimensional codewords~\cite{GPU-DPF}:
\[
\{C_0\in \mathbb{F}_{2^\lambda}^{2\times(log(N)+1)}, C_1\in \mathbb{F}_{2^\lambda}^{2\times(log(N)+1)}\}
\]
These keys are then sent to the respective servers~\ding{203}.


\subsection{DPF evaluation}
In the DPF construction of \cite{DPF}, DPF evaluation is based on a GGM PSF computation tree, a binary tree where pseudorandom values are expanded at each level, starting from a root seed.
The goal is to obtain a vector $\mathbf{v} = [\mathsf{Eval}(k, 0), \mathsf{Eval}(k, 1), \dots \mathsf{Eval}(k, N-1)]$ where $k \in \{k_1, k_2\}$.
To evaluate the function shares: $\mathsf{Eval}(k,\cdot)$, the GGM PSF tree uses a recursive function $R$ as follows:

\begin{align*}
 \mathsf{Eval}(k,j) = R(d=log(N), j)\tag{1}\\
 R(0,0)=C_0[0,0]\tag{2}\\
 R(d,j) = \mathsf{PRF}_{R(d-1,\lfloor \frac{j}{2} \rfloor )}(j~mod~2) +\\ C_{R(d-1,\lfloor \frac{j}{2} \rfloor)~mod~2}[j~mod~2, d] \tag{3}
\end{align*}
$d$ represents the depth of a node in the tree: $0$ for the root node and $log(N)$ for the leaf nodes; $j$ is the index of the node within each depth(0 being the leftmost index); and $\mathsf{PRF}_s(x)$ is a pseudorandom function which encrypts a value $x$ with an encryption key $s$.
A commonly used PSF (also used in this work) is AES-128.
For example, to query a database containing four items, each server computes: $\mathsf{Eval}(k,0), \mathsf{Eval}(k,1), \mathsf{Eval}(k, 2), \mathsf{Eval}(k,3)$ for their respective keys.
Following from (1): 
\[
\mathsf{Eval}(k,j) = R(d=log(N), j) = R(log(4),j) = R(2,j)
\]
This means the servers must compute: $R(2,0), R(2,1), R(2, 2)$ and $R(2,3)$.
\begin{figure} [!t]
	\centering
	\includegraphics[width=1\linewidth]{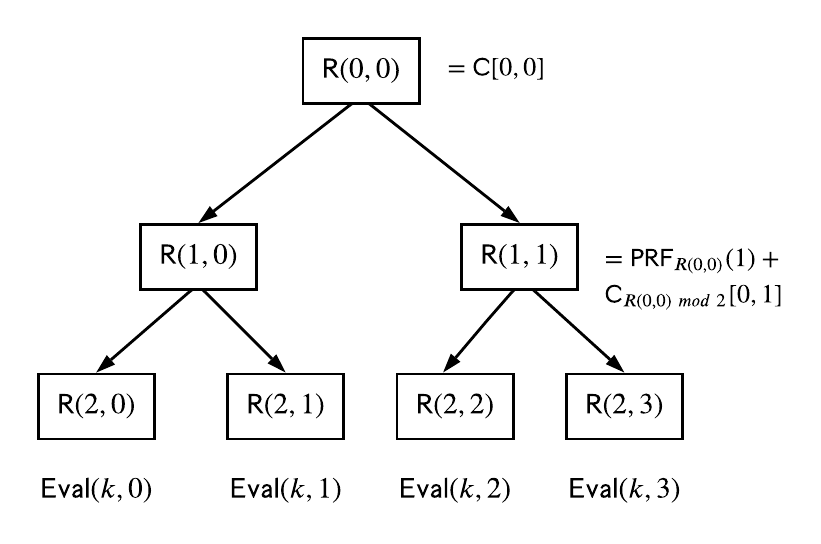}
	\caption{GGM computation tree for $N=4$. Each node invokes a PRF, seeded with values from previous nodes. The leaf nodes correspond to $\mathsf{Eval}(k,j)$.}
	\label{fig:eval-bin-tree}
\end{figure}
\autoref{fig:eval-bin-tree} illustrates the binary tree construction needed for computing these values.

\subpoint{DPF Eval parallelization.}
\begin{figure} [!t]
	\centering
	\includegraphics[width=1\linewidth]{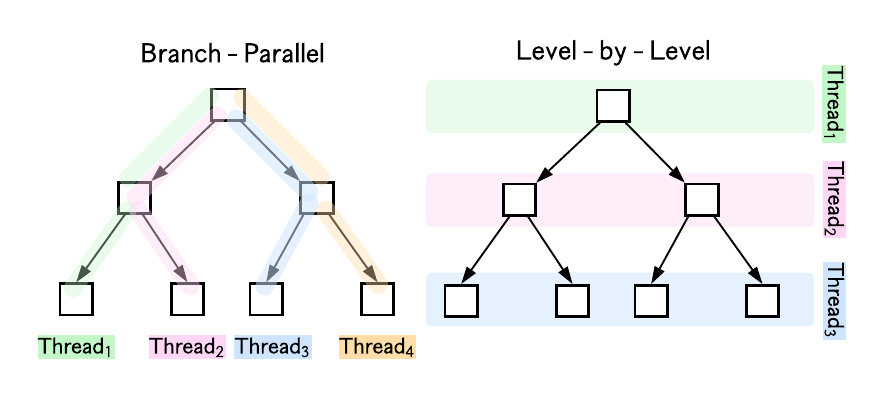}
	\caption{Common parallelization approaches for DPF evaluation.}
	\label{fig:eval-parallel}
\end{figure}
On preliminary investigation, we observe that the DPF evaluation tree is amenable to parallelization.
Two common approaches for proceeding are a \emph{branch-parallel} and a \emph{level-by-level} approach, illustrated in \autoref{fig:eval-parallel}.
In the branch-parallel approach, each thread is responsible for computing one or more leaf nodes, \ie $\mathsf{Eval}(k,j)$), in the DPF evaluation tree.
However, in this approach, each thread redundantly performs the full path from the root node to the leaf, leading to wasteful computations.
On the other hand, in the level-by-level approach, computation threads are assigned to tree levels (\ie tree depths), with each thread computing all nodes at a single level and saving the intermediate results. 
While this approach eliminates the redundant path computations in the branch-parallel approach, it requires extensive synchronization between threads at each level, and consumes a large amount of memory for storing intermediate results for each level.
A previous study on GPU-based DPF evaluation~\cite{GPU-DPF} provides an extensive discussion of the tradeoffs between branch-parallel and level-by-level approaches for computing the DPF evaluation tree.
They proceed with a more efficient technique called memory-bounded tree traversal, which is similar to the level-by-level approach, but computes smaller chunks of nodes per level, rather than the full level.

We discuss these parallelization techniques from a UPMEM PIM DPU perspective.

\subsubpoint{Branch-parallel.}
Although this approach introduces computational redundancy, it remains the most intuitive method and can still be considered for PIM-based processing by leveraging the massive parallelism offered by UPMEM PIM DPUs.
Nevertheless, architectural details of UPMEM PIM, specifically the limited WRAM size of DPUs ($64$\,KB), make this approach infeasible on the current UPMEM PIM architecture.

\subsubpoint{Level-by-level.}
As regards the level-by-level and memory-bounded tree traversal approaches, they both require significant inter-DPU communication to share intermediate results among DPUs.
Unfortunately, the current UPMEM PIM architecture does not support direct DPU-DPU communication, and all inter-DPU data copying must go through the host CPU.
Several studies~\cite{gilbert24, PIM_HE, friesel23, hyun24} discuss the overhead of inter-DPU communication, showing how its cost alone can largely outweigh any benefits obtained from offloading these highly parallelizable operations on DPUs.

Lastly, in the GGM-based DPF evaluation, each node invokes a pseudorandom function, AES-128 in this case.
PSFs like AES-128 involve multiple rounds of compute-intensive operations not suitable for lightweight cores like 32-bit RISC DPUs.
Moreover, unlike CPUs which feature dedicated cryptographic instructions such as AES-NI~\cite{AES-NI}, UPMEM PIM DPUs lack any crypto acceleration primitives, making them unsuitable for state-of-the-art PSFs.

These points suggest that PIM DPUs are not good options for offloading the tree-based operations in DPF evaluation.
As a result, \sys offloads this work to the host-side CPU, which performs DPF evaluations~\ding{203} using cryptographic hardware-accelerated AES-NI instructions. 

To parallelize the DPF evaluations, we partition the DPF evaluation tree into subtrees (each a perfect binary tree\footnote{A perfect binary tree is one in which every level is completely filled and all leaf nodes are at the same level.}) and assign each subtree to a CPU worker thread.
For example, suppose we have $N_t$ CPU worker threads, where $N_t$ is a power of two: 2, 4, 8, etc.
We choose a (non-leaf) level L of the DPF tree such that $N_t = 2^L$.
The nodes at this level are roots of $N_t$ (perfect) binary trees, which are then evaluated in parallel by the $N_t$ worker threads.
A master thread performs the DPF evaluation (following a breadth first traversal approach) from the root down to level L.
Each worker thread then takes over and completes DPF evaluation on its subtree following a similar breadth first traversal approach to obtain a subset of the $\mathsf{Eval}(k,\cdot)$ function shares.

\subsubpoint{AES-NI optimization.}
Cryptographic hardware instructions like AES-NI provide a pipelined approach enabling execution of multiple AES operations in parallel.
At each level of a subtree, we batch AES calls across multiple nodes to maximize utilization of this pipeline.

Once the host CPU computes the function shares \ie $\mathsf{Eval}(k,\cdot)$, these values are copied in chunks to the DPUs~\ding{204}.
In the following, we outline how this chunking is done.

\subsection{PIR linear operations: $\mathsf{dpXOR}$}
Here, we explain how data is distributed across the DPUs, and how the latter perform linear operations ($\mathsf{dpXOR}$) to produce the subresults required for obtaining the desired DB item client-side.

\subpoint{Database preloading.}
As previously discussed, efficiently offloading computations to DPUs hinges on minimizing data transfer (CPU-DPU and DPU-DPU) overhead; excessive data movement can undermine the benefits of in-memory processing. 
Consequently, the performance gains of the PIR acceleration algorithms rely on reducing both the frequency and volume of data transfers between the host CPU and the DPUs.

To perform PIM-based PIR evaluations, both the database and the bit arrays (\mbox{\ie} secret shares generated via the DPF evaluation) must reside in DPU memory during execution. 
Because the database is typically large, frequent CPU--DPU transfers would incur significant overhead. 
To mitigate this, \mbox{\sys} preloads the database into DPU MRAM before query processing. The bit arrays, in contrast, are query-dependent and are supplied to the DPUs dynamically at execution time.

In our setup, the database is loaded following a linear (one-dimensional) layout, with each DPU's MRAM accommodating a chunk/block $B_d$ of the database items, with $B_d = \lceil N/P \rceil$, where $P$ is the number of PIM DPUs. 
For example, for a system with $2048$ DPUs and database containing $N=2^{32}$ one byte items, we have: $B_d=\lceil 2^{32}/2048 \rceil = 2^{21}$ database items per DPU.
So for $D = [d_0, d_1, \dots, d_N]$, the first DPU gets the DB chunk $[d_0, \dots, d_{B_d-1}]$, the second DPU gets DB chunk $[d_{B_d}, \dots, d_{2B_d-1}]$ and so on.
Each database block/chunk is then copied to a DPU.
The results from the DPF evaluations in the previous step are distributed across the DPUs in the same fashion as database chunks.
So the first DPU receives the first $B_d$ DPF evaluation results:  $\{Eval(k,0),\dots ,Eval(k,B_d-1)\}$, the second DPU $\{Eval(k,B_d),\dots ,Eval(k,2B_d-1)\}$ and so on.

One key consideration with this approach is the memory capacity of PIM-enabled hardware. 
Current UPMEM PIM-enabled servers support up to 20 UPMEM modules (expected to increase in the future).
This configuration provides up to $160$\,GB of MRAM, which is sufficient to accommodate mid- to large-scale PIR databases. 
Larger datasets may require a minor adaptation of our "one-shot" (\ie single pass) database evaluation: for example, by evaluating the linear operations on database items in batches, copying unprocessed chunks into DPUs in each batch. 
Nevertheless, as PIM technology continues to evolve, the memory capacities and capabilities are expected to expand, opening new possibilities for efficient and scalable data processing.
For frequently updated databases, DPUs can handle queries on a stable version of the database, while the CPU uses brief windows when DPUs are idle to apply bulk database updates; this amortizes the CPU--DPU transfers and synchronization overheads.

\subpoint{Inner products and XORing operations.}
The goal of this stage is to compute a linear combination of the database entries weighted by the query vector (represented by the function shares $\mathsf{Eval}(k,\cdot)$) so as to obtain two subresults: $r_1$ and $r_2$ at both servers 1 and 2 respectively~\ding{205}.
To obtain a subresult $r$ (\ie  $r_1$ or $r_2$), a server proceeds by doing:
\[
r = \bigoplus_{j=0}^{N-1}\mathsf{Eval}(k,j)\cdot D[j]
\] 
Since the database is distributed across $P$ DPUs, each DPU performs only a part of the above operation to obtain a subresult $s_d$ where $d \in \{0, P-1\}$.
Assuming the start and end index of a DPU's database block are respectively $dstart$ and $dend$, the DPU computes $s_d$ as follows:
\[
s_d = \bigoplus_{j=dstart}^{dend}\mathsf{Eval}(k,j)\cdot D[j]
\] 

We exploit the PIM system's inherent two-tiered parallelism by utilizing both multiple DPUs and multiple DPU threads, \ie tasklets, to perform PIR operations within each DPU. 

\subsubpoint{Parallel reduction (PR).}
Within each DPU, the workload is further partitioned among tasklets, and a two-stage \emph{parallel reduction} strategy is employed to perform the XORing operations on each DPU's database chunk $B_d$.
Parallel reduction~\cite{harris2007optimizing} is a divide-and-conquer technique where a large dataset is processed concurrently by multiple threads to derive a single result.
In the first stage of parallel reduction, the MRAM-resident database chunk $B_d$ is split among the DPU's tasklets such that each tasklet processes $B_t = \lceil B_d/T \rceil$ items, where $T$ represents the number of tasklets per DPU.
Similarly, the subset of $B_d$ $\mathsf{Eval}(k,j)$ items from the DPF evaluation copied to each DPU is equally split among the DPU's tasklets, with each obtaining $B_t$ entries.  
Each tasklet $T_i$ then computes a partial result, $t_i$ by XORing the relevant DB items.
The DB items included in the XOR result are determined by the corresponding DPF evaluation result, which acts as a selector for the DB item: if $\mathsf{Eval}(k,j)$ bit is set (1), then $D[j]$ is included in the XORed result, otherwise $D[j]$ is ignored.
A master tasklet (tasklet 0) waits until the $T$ assigned tasklets have computed their partial results, and in the second stage of parallel reduction, this master tasklet then XORs the partial results from all tasklets to obtain a final subresult $s_i$ for DPU-i.

Once computed, the subresults are copied from each DPU back to the host~\ding{206}, and aggregated to obtain the server's subresult $r = \bigoplus_{d=0}^{P-1}s_d$~\ding{207}.
Each server's subresult is sent back to the client who combines the subresults $r_1$ and $r_2$~\ding{208} to retrieve the requested DB item $D[i] = r_1 \oplus r_2$.

The end-to-end operation of \sys, beginning with the client's key/query generation and ending with result retrieval, is summarized in \autoref{alg:impir-operation}.

\begin{algorithm}[th!]	
	\caption{End-to-end operation of \sys.}
	\begin{algorithmic}[1]
		\State \textbf{Global variables:}
		\State \text{P}: num. of DPUs
		\State \text{$B_d$}: num. of DB items per DPU
		\State \text{T}: num. of DPU tasklets
		\State \text{$B_t$}: num. of DB items per tasklet		
		\State \underline{\textbf{Client:}} \Comment{\textcolor{gray}{Executes on the client}}
		\Procedure{\textsc{GenerateAndSendKeys}}{\textit{i}}~\ding{202}
			\State	$(k_1, k_2) \gets \mathsf{Gen}(i, 1)$
			\State \text{Send $k_1$ to $\mathsf{Server_1}$ and $k_2$ to $\mathsf{Server_2}$}
		\EndProcedure
		
		\\
		\State \underline{\textbf{Host:}} \Comment{\textcolor{gray}{Executes on the host CPU of each DB server}}
		
		\Function{EvaluateDPF}{\textit{$k \in \{k_1, k_2\}$}}~\ding{203}
			\For{$j=0$ \textit{to} $N-1$}
				\State $\mathsf{Eval}(k, j) \gets \mathsf{R}(d=log(N), j)$ \Comment{\textcolor{gray}{Uses AES-NI}}
			\EndFor	
			\State \Return $\mathbf{v} \gets [\mathsf{Eval}(k, 0), \dots, \mathsf{Eval}(k, N-1)]$
		\EndFunction
		
		\Procedure{SplitDPF}{\textit{$[\mathsf{Eval}(k, 0), \dots, \mathsf{Eval}(k, N-1)]$}}~\ding{204}
			\For{$i=0$ \textit{to} $P-1$}
				\State $\mathsf{chunk_i} \gets [\mathsf{Eval}(k, i*B_d), \dots, \mathsf{Eval}(k, i*B_d + B_d-1)]$
				\State $\mathsf{Copy}(\mathsf{chunk_i})$ \textit{to} $\mathsf{DPU_i}$
			\EndFor			
		\EndProcedure
		
		\State \underline{\textbf{DPU:}} \Comment{\textcolor{gray}{Executes on a PIM DPU of the DB server}}
		\State $D_d = [d_{dstart}, \dots, d_{dend}]$ \Comment{\textcolor{gray}{Preloaded DPU DB block}}
		\State $\mathbf{v} = [\mathsf{Eval}(k, dstart), \dots, \mathsf{Eval}(k, dend)]$
		\Function{TaskletXOR}{\textit{$\mathbf{v}, D_d$}}~\ding{205} \Comment{\textcolor{gray}{PR: stage 1}}	
			\State \Comment{\textcolor{gray}{All tasklets execute this function on parts of the DB.}}
			\State $t_i \gets 0$ \Comment{\textcolor{gray}{Tasklet's partial result}}
			\State $tid \gets $ \textsc{getTaskletID()}\Comment{\textcolor{gray}{$tid \in \{0, \dots, T-1\}$}}
			\For{$i=tid*B_t$ \textit{to} $tid*B_t + B_t - 1$} 
				\If{$\mathbf{v}[j] = 1$}
					\State $t_i \gets t_i \oplus D_d[j]$
				\EndIf		
			\EndFor	
			
			\State \Return $t_i$ \Comment{\textcolor{gray}{All $t_i$s will produce $\mathbf{t} = [t_0, \dots, t_{T-1}]$}}
		\EndFunction
		
		\Function{MasterXOR}{\textit{$\mathbf{t} = [t_0, \dots, t_{T-1}]$}}~\ding{205} \Comment{\textcolor{gray}{PR: stage 2}}
			\State $s \gets 0$ 			
			\For{$i=0$ \textit{to} $T-1$} 
				\State $s \gets s \oplus \mathbf{t}[i]$	
			\EndFor	
			\State \Return $s$
		\EndFunction
		\State $s = [s_1, \dots, s_{P-1}] \gets \mathsf{Copy}(subresults)$ \textit{from P DPUs}~\ding{206}  			
		\Procedure{AggregateSubresults}{$[s_1, \dots, s_{P-1}]$}			
			\State $r \gets 0$ 			
			\For{$i=0$ \textit{to} $P-1$}~\ding{207}			
				\State $r \gets r \oplus s_i$	
			\EndFor
			\State \text{Send $r \in \{r_1, r_2\}$ to $\mathsf{Client}$}	
		\EndProcedure
		\State \underline{\textbf{Client:}} 
		\State $D[i] = r_1 \oplus r_2$~\ding{208}\Comment{\textcolor{gray}{Reconstruct DB item}}
	\end{algorithmic}
	\label{alg:impir-operation}
\end{algorithm}

\subsection{Batching client queries} 
\begin{figure}[th!]
	\centering
	\includegraphics[width=1\linewidth]{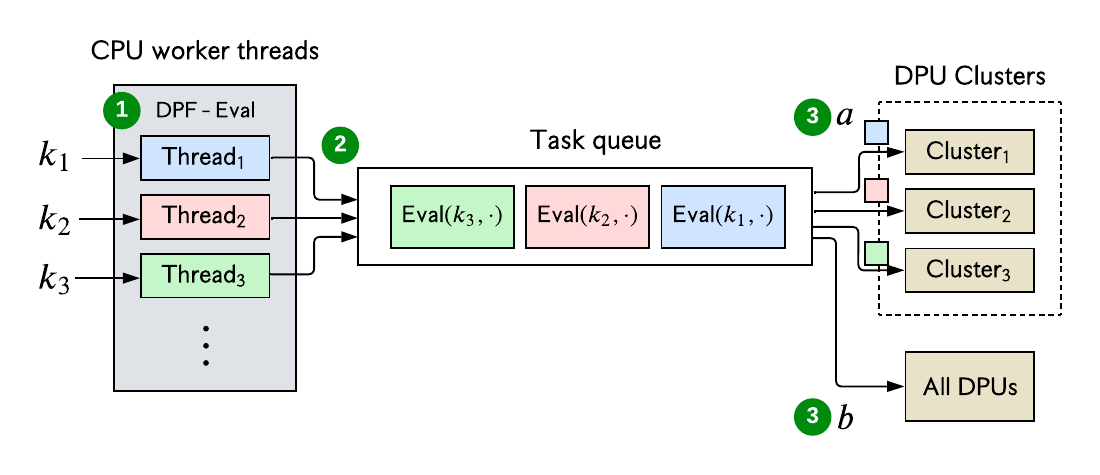}
	\caption{Workflow for multi-query processing with \sys: worker threads compute per-key DPF evaluations and add their results to a task queue. Dedicated threads assign queries, represented by $\mathsf{Eval}(k,\cdot)$ function shares to specific DPU clusters for processing, and aggregate results as described in \autoref{alg:impir-operation}.}
	\label{fig:batch_execution}
\end{figure}

A PIR server often handles multiple queries simultaneously, \ie query batching; this requires an efficient strategy to maximize query throughput.
In our multi-server PIR approach, this translates to sending multiple keys (instead of a single one) to each DB server, and the latter handling the queries in a concurrent fashion.
To support concurrent processing of multiple queries, \sys partitions the host- and PIM-side computational resources, \ie CPU and DPUs respectively, across the different queries; this is summarized in \autoref{fig:batch_execution}
On the host CPU side, a set of $W$ worker threads are created to simultaneously manage DPF evaluations for the keys (\ie queries) received from the client~\textcolor{pimcolor}{\ding{202}}.
The keys are split across these worker threads which perform DPF evaluations for the received keys.
On the PIM-side, the DPUs are organized into clusters of $P_c$ DPUs, each cluster handling a single PIR query (\ie key).
As a host-side worker thread completes DPF evaluations for a key $k_i$, it adds the result $\mathbf{v_i} = [\mathsf{Eval}(k_i,0),\dots, \mathsf{Eval}(k_i,N-1)$ to a shared task queue~\textcolor{pimcolor}{\ding{203}}.
\sys dedicates some host-side CPU threads for DPU management: these threads are responsible for fetching tasks from the task queue and scheduling DPU clusters to handle the associated queries in parallel~\textcolor{pimcolor}{\ding{204}}-a.
The DPU clusters perform the $\mathsf{dpXOR}$ operations as described in \autoref{alg:impir-operation}~\ding{205}-\ding{206}, and the scheduling threads aggregate the partial results from the clusters following \autoref{alg:impir-operation}~\ding{207}.
Also, queries can be handled sequentially such that all DPUs are deployed for a single query~\textcolor{pimcolor}{\ding{204}}-b.
The choice between sequential query processing (using all DPUs) or parallel query processing on multiple smaller DPU clusters simultaneously depends on factors such as the size of the database and the number of queries.
For very large databases, the sequential strategy could be preferred as this allows for in-place processing on the entire DB, rather than on chunks of the DB (if all $P_c$ DPUs cannot accommodate the entire DB).
For smaller databases where a smaller DPU cluster with $P_c$ DPUs can accommodate the full DB, the clustered approach will be preferred. 
\section{Implementation}
\label{sec:implem}
\sys is implemented on UPMEM PIM, the first openly commercialized PIM architecture.
The client-side and server-side code ($\approx 2500$\,LoC) are implemented in C++, except for the DPU code ($\approx 200$\,LoC) which is implemented in C.
The DPF implementation use AES-128 as its pseudorandom function (PRF).
On the host CPU of the PIR server, we leverage Intel AES-NI cryptographic extensions and advanced vector extensions (AVX) for 256-bit operations to accelerate DPF evaluation.

\section{Evaluation}
\label{sec:evaluation}
In this section, we describe our experimental evaluation of \sys and compare its performance against traditional processor-centric PIR implementations: CPU- and GPU-based.
The CPU-based PIR (our baseline) is a DPF-PIR implementation from Google Research~\cite{google-dpf-pir}, while the GPU-based PIR is prior work by Meta AI and Harvard~\cite{GPU-DPF}.

We aim to answer the following questions:
\begin{itemize}[]
	\item[\textbf{Q1}:] How does \sys improve PIR query latency and throughput compared to a traditional CPU-based PIR baseline? (\S\ref{eval:pir-tput})
	\item[\textbf{Q2}:] What are the benefits of \sys's DPU clustering strategy for batch execution of queries? (\S\ref{eval:cluster})
	\item[\textbf{Q3}:] How does \sys compare against related work~\cite{GPU-DPF} on GPU-based PIR acceleration? (\S\ref{eval:gpu} )
\end{itemize}


\subsection{Evaluation methodology}
We evaluate the performance of \sys across multiple dimensions, including latency, throughput (queries per second), and scalability using a real UPMEM PIM server with a host side CPU and PIM DPUs.
We consider a CPU-based multi-server PIR implementation as our baseline.
Because the PIM server's PIM-equipped DIMMs occupy memory channels that could otherwise have been occupied by standard DRAM, it is unfair to run the baseline CPU PIR implementation on this server, as the presence of PIM memory limits the total potential (DRAM) memory bandwidth of the server.
For this reason, we run the CPU PIR baseline on a separate server without any PIM support.
This "dual-server" setup for fairness is well-founded, as shown in past studies~\cite{PIM-TREE, pim-join, gomez22} which compare PIM-based solutions to their purely CPU-based counterparts on separate servers.
The CPU PIR baseline uses a single CPU thread for each query, but leverages Intel AVX for acceleration.

In our PIM-based PIR implementation, the PIR database is preloaded once into PIM MRAM memory.
As such data transfer time for this data is not part of our evaluations.
On the other hand, we discuss data transfer times for DPF function shares as well as DPU subresults, since these are part of every PIR query evaluation.

Our experimental results include only PIR server-side evaluations, as our contribution, \sys, is entirely server-side.
As such, our results do not include time for key generation or final result reconstruction at the client; these are trivial low-latency operations which do not require PIM-based acceleration.
Similarly, client-server communication latency is not included since \sys has no effect on these.
Since the server-side evaluations are (theoretically) identical for all the PIR database servers in a multi-server PIR setup, our results simply focus on one PIR database server.

Unless otherwise stated, we report the mean over 10 runs.

\subsection{Evaluation setup}
\subpoint{UPMEM PIM platform.}
We evaluate \sys on a UPMEM PIM enabled server equipped with 20 PIM-enabled memory modules, totaling 2560 DPUs, each operating at $350$\,MHz.
This represents a total of $160$\,GB of MRAM memory.
The actual memory bandwidth between each DPU and its associated MRAM bank is $\approx700$\,MB/s, resulting in an aggregate memory bandwidth of $\approx1.79$\,TB/s. 
The PIM server has two 8-core Intel(R) Xeon(R) Silver 4110 CPUs clocked at $2.10$GHz with hyper-threading enabled; each CPU has a last-level cache of size 11\,MB.
The CPUs support AVX for 256 bit operations.
The server equally has $256$GB of standard DRAM.

We configure each DPU to run 16 tasklets concurrently, which is an acceptable number of tasklets to fully utilize the DPU pipeline (above 11 tasklets is recommended~\cite{PRIM,upmem}).
Unless stated otherwise, our experiments use a total of $2048$ DPUs.\footnote{It is usually easier to work with powers of 2.}

\subpoint{Traditional machine w/o PIM.}
This server is used to run the standard CPU-based PIR design, which is our baseline PIR approach.
The server is equipped with two 16-core Intel(R) Xeon(R) E5-2683 v4 CPUs clocked at $2.10$\,GHz with hyper-threading enabled; each CPU has a last-level cache of size 40\,MB.
The server has $128$\,GB of standard DRAM memory.

\subpoint{GPU platform.}
We run all GPU-based experiments on an NVIDIA GeForce RTX 4090 GPU operating at $2235$\,MHz with a last-level cache of size $72$\,MB.
It has a memory (VRAM) size of $24$\,GB and a total memory bandwidth of 1.01\,TB/s.

\subpoint{PIR database.} 
To emulate the behavior of real PIR databases/use cases, we generate a PIR database where each record is a random records 32-byte (256-bit) hash. 
This data format is widely used across various security- and integrity-critical applications.
For example, Certificate Transparency (CT) auditing~\cite{simplePIR, ypir} uses SHA-256 to store records of issued SSL/TLS certificates, allowing organizations to detect fraudulent or misissued certificates~\cite{GoogleCT,ChromiumCT}. 
Also, compromised credential verification services, such as \textit{Have I Been Pwned}~\cite{HIBP} and enterprise password managers, use SHA-256 hashes to store and compare leaked password hashes against user-submitted credentials~\cite{HIBP,NISTPasswords,OWASPPasswords}.


\subsection{Impact of \sys on PIR query throughput and latency}
\label{eval:pir-tput}
\begin{figure}[!t]
	\centering
	\includegraphics[width=1\linewidth]{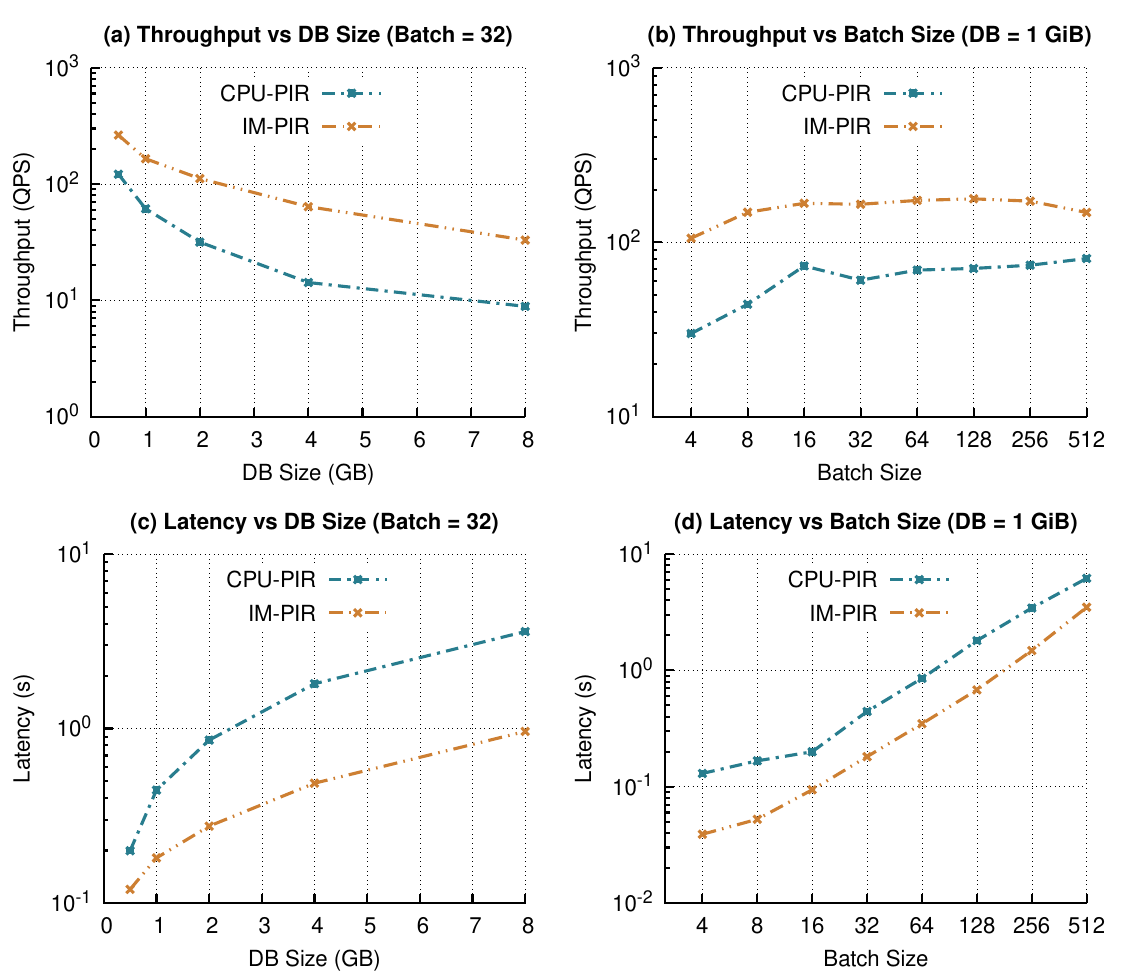}
	\caption{Comparison of query throughput and latency between \sys and a CPU-based PIR design across different database sizes and number of queries (batch sizes). Each database record is $32$ bytes in size. The CPU version is accelerated with AES-NI and AVX instructions and has each query handled by a CPU thread. The plots are in log scale.}
	\label{fig:throughput}
\end{figure}

Here, we evaluate the impact of our PIM-based PIR design (using different numbers of DPUs per query) on query throughput (number of queries per second, QPS) and latency (total computation time for all queries), and compare these against a CPU-based approach (henceforth CPU-PIR).
\autoref{fig:throughput} illustrates the results.

\subpoint{Throughput/latency with varying DB sizes.}
In \autoref{fig:throughput} (a) and \autoref{fig:throughput} (c), we maintain a batch of 32 queries (encoded as PIR keys) and show the resulting throughputs and latencies respectively, after performing these queries to completion on the PIR server for varying DB sizes.
That is, from DPF evaluation using the input keys to aggregation of DPU subresults, \ie \autoref{alg:impir-operation}~\ding{203} to \ding{207}.
For small databases (\ie 0.5\,GB), the query throughput in \sys is $1.7\times$ larger when compared to the CPU-based PIR baseline.
While the actual throughput decreases with larger DB sizes, the overall throughput improvement with \sys increases to over $3.7\times$ for a DB size of 8\,GB.
The throughput improvement in \sys is mainly as a result of the massively parallel in-place processing of the PIR database, in contrast to the CPU-based approach where the DB items must be moved from memory to the CPU for processing.
As the DB size increases, the advantage of \sys becomes more apparent since CPU-PIR suffers from extensive data movement overhead between DRAM and the CPU.

We note that in this experiment, a single DPU cluster is used (see \autoref{fig:batch_execution}~\textcolor{pimcolor}{\ding{204}}-b), while the maximum number of threads (32) is used in CPU-PIR. 
As such, in \sys, all the DPUs are used for each query in the batch, meaning each query (specifically the \dpxor operations) is processed sequentially.
This, unfortunately, leads to limited parallelism and does not showcase the full potential of \sys. 
We defaulted to this setup so as to accommodate larger DB sizes.
However, with smaller DB sizes (or larger number of DPUs), the queries's \dpxor operations can all be handled in parallel by creating DPU clusters which hold the entire DB and can each process a query concurrently.
This will lead to an even larger throughput improvement in \sys with respect to CPU-PIR.
We showcase the advantage of this clustering approach in \S\ref{eval:cluster}.


\observ{\sys significantly improves query throughput, by up to $3.7\times$ compared to the baseline CPU PIR design. This is due to the massively parallel in-place processing of the PIR database in \sys.}

From \autoref{fig:throughput} (c), we observe that for both \sys and CPU-PIR, query latency increases linearly with database size, which is expected.
However, \sys scales much better (the slope is smaller) compared to CPU-PIR.
As the database size increases, CPU-PIR suffers more cache misses as its last-level cache cannot accommodate the large DB; hence performance degrades more sharply.
In contrast, \sys can process the entire DB in-place, leading to much better performance. 

\subpoint{Speedup factor.} 
To better quantify the performance gain of \sys relative to CPU-PIR, we define the speedup factor as the ratio of CPU-PIR query latency to \sys query latency.  
We observe that PIM achieves a speedup of $\approx 2\times$ at 0.5\,GB, which increases to over $3.7\times$ at 8\,GB. 
This trend suggests that larger database sizes aggravate the CPU’s memory bandwidth limitations, while DPUs, despite having relatively limited computation ability, provide improved performance due to the capability of in-place DB processing with minimal data movement.
Similarly, we note that these results are achieved while running multiple queries on a single DPU cluster, \ie using all DPUs for each query.
For smaller databases, \eg 1\,GB, \sys has potential for even larger speedup with respect to (wrt.) CPU-PIR as queries may be processed in parallel across the DPU clusters.
These results indicate multi-server PIR operations are indeed better adapted to a PIM execution model, as opposed to a CPU-centric model.

\observ{\sys provides up to $3.7\times$ latency speedup wrt. a CPU-centric multi-server PIR design. This performance gain is mainly due to the ability to process large databases in-place via PIM, as opposed to a CPU-centric design where the relatively small last-level cache cannot hold the entire DB, leading to costly data movement between memory and CPU.}

\subpoint{Throughput/latency with varying batch sizes.}
In \autoref{fig:throughput} (b) and \autoref{fig:throughput} (d), we fix the DB size at 1\,GB while varying the query batch size, and show the resulting throughputs and latencies (respectively) after evaluating the queries to completion on the database.
As the query batch size increases, \sys's throughput remains fairly constant, but is $2.6\times$ higher on average wrt. CPU-PIR.
As explained previously, this experiment uses a single DPU cluster with all DPUs for processing each query, which explains the fairly constant throughput across batch sizes.
Nevertheless, with more DPU clusters, the overall throughput is expected to increase as more queries will be processed in parallel over the DPUs (see \S\ref{eval:cluster}).
This means a much larger throughput improvement is expected compared to the CPU-based PIR approach.
Similarly, the query latencies in \sys and CPU-PIR increase linearly with batch size as more DPF evaluations and \dpxor operations are done in total.
But \sys has overall lower latencies due to in-place DB processing, as explained before.

%
%

\begin{figure}[!t]
	\centering
	\includegraphics[width=1\linewidth]{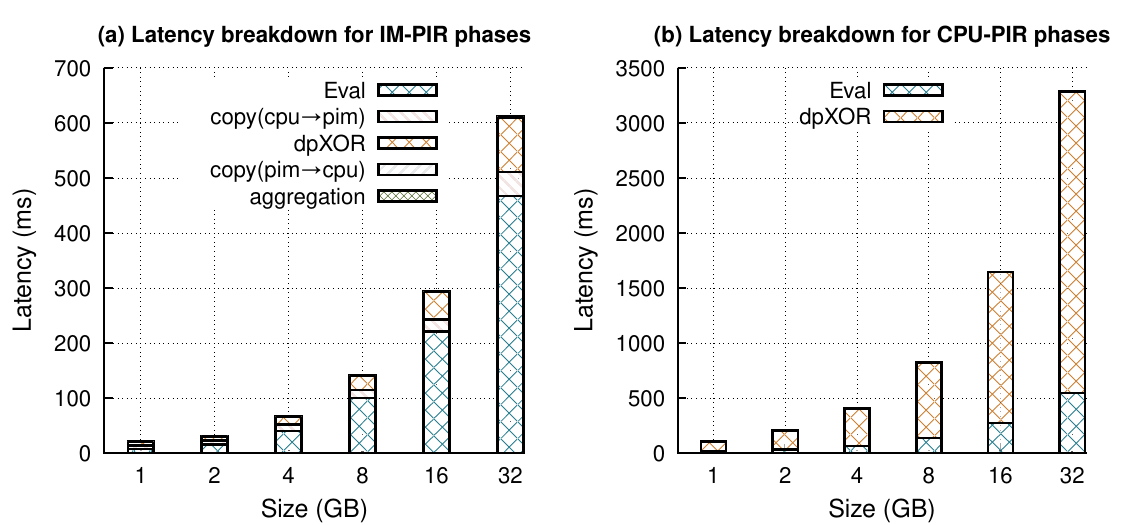}
	\caption{Latency breakdown showing cost of all server-side phases/steps in \sys and CPU-PIR. In CPU-PIR, the memory bandwidth limitations make \dpxor operations on the DB the primary bottleneck. In \sys, in-place DB processing significantly improves the memory-bound \dpxor operations, making the CPU-side DPF evaluation the primary bottleneck.}
	\label{fig:component_analysis}
\end{figure}

\subpoint{Latency breakdown for \sys query execution phases.}
Here, we perform a breakdown of \sys's query execution time into the following phases: DPF evaluation, data transfer from CPU to DPU of function shares $\mathsf{Eval(k,\cdot)}$, \dpxor operations on DB items, copying of subresults from DPU to CPU, and finally aggregation of these subresults on the host CPU; these phases correspond to \autoref{fig:pir-pim} (and \autoref{alg:impir-operation}) steps \ding{203}, \ding{204}, \ding{205}, \ding{206}, and \ding{207}, respectively.
This component-wise analysis provides important insights regarding the primary performance bottlenecks in our approach.
\autoref{fig:component_analysis} illustrates the results of this component-wise analysis for \sys and a CPU-based PIR approach and \autoref{tb:latency-breakdown} summarizes the average percentage contributions of each phase

\renewcommand{\arraystretch}{1.5}
\begin{table}[!t]
	\centering
	\resizebox{\columnwidth}{!}{ %
		\begin{tabular}{lccccc}
			\toprule
			\textbf{PIR approach} & \textbf{DPF Eval} & \textbf{CPU$\rightarrow$DPU copy} & $\textbf{dpXOR}$ & \textbf{DPU$\rightarrow$CPU copy} & \textbf{Aggreg.}\\
			\midrule
			\sys & 76.45\% & 7.17\% & 16.20\% & 0.18\% & 0.00002\% \\
			CPU-PIR& 16.64\% & N/A & 83.36\% & N/A & N/A \\
								
			\bottomrule
			\rowcolor{white}
		\end{tabular}
	}
	\caption{Average percentage contribution to overall query latency of server-side PIR execution phases for \sys and CPU-PIR.}
	\label{tb:latency-breakdown}
	\vspace{-10pt}
\end{table}

We observe that in CPU-PIR, the primary performance bottleneck is \dpxor ($83.36\%$ of total query execution time), whereas the primary performance bottleneck in \sys is DPF evaluation operations ($76.45\%$ percent of total query execution time).
In CPU-PIR, the \dpxor operations are highly constrained by the limited memory bandwidth, as the entire DB needs to be processed for each query.
\sys addresses this memory bandwidth limitation, providing extensive scaling of compute resources with memory, thus resulting in improved performance for the DB \dpxor operations (only $16.20\%$ of total query execution time).
Meanwhile, the compute-intensive DPF evaluation becomes the primary bottleneck.
Data copy overhead between the CPU and DPUs contributes to less than $8\%$ of the total cost because the the DPF function shares (bit vectors) and DPU subresults are relatively lightweight.
Moreover, the DB is preloaded in DPU MRAM, and thus DB copy overhead has no impact on query latency.

\observ{In a CPU-based PIR, the \dpxor operations on the database constitute the primary performance bottleneck. However, in a PIM-based approach, the extensive parallelism and in-place \dpxor operations on the in-memory DB significantly accelerates the \dpxor operations. As a result, the host-side CPU-based DPF evaluation becomes the primary PIR bottleneck.}

\subsection{DPU clustering}
\label{eval:cluster}
\begin{figure}[!t]
	\centering
	\includegraphics[width=1\linewidth]{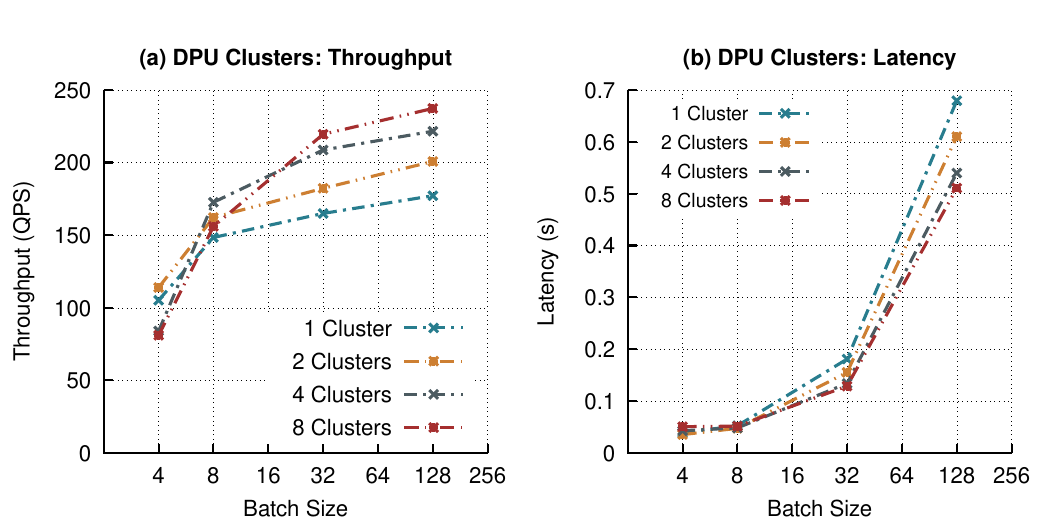}
	\caption{Effect of DPU clustering on query throughput and latency in \sys for different query batch sizes. For a single cluster, all 2048 DPUs are used; for two clusters, each cluster has $2048/2 = 1024$ DPUs, \etc. The database size is kept constant a 1\,GB.}
	\label{fig:dpu_clustering}
\end{figure}

Here, we illustrate the effect of \sys's DPU clustering approach on query throughput and latency for different query batch sizes, while maintaining the DB size at 1\,GB.
One cluster means a total of 2048 (\ie all) DPUs; with two clusters, the DPUs are evenly split across all clusters, resulting in $2048/2 = 1024$ DPUs per cluster, and so forth.
\autoref{fig:dpu_clustering} illustrates the results.

We observe that for a given batch size, overall query throughput increases with a larger number of DPU clusters; with up to $1.35\times$ throughput improvement with 8 DPU clusters compared to a single cluster.
Conversely, the smallest number of clusters where all DPUs are used presents the lowest throughput in general.
This is due to the fact that larger numbers of clusters allows for more queries to be processed in parallel.
That is, once host-side CPU threads in \sys perform DPF evaluations on keys and add their results ($\mathsf{Eval(k,\cdot)}$) to the task queue (see \autoref{fig:batch_execution}), each DPU cluster can perform \dpxor operations over the entire database.
This improved parallelism leads to better throughput.
In the single cluster approach, each query's \dpxor operations must be performed serially: \ie all \dpxor operations from one query must be completed before the next one is processed, resulting in poorer performance.
Similarly, as shown in \autoref{fig:dpu_clustering} (b), we have lower overall query latency for larger DPU clusters due to the improved parallelism.

\observ{DPU clustering can improve query throughput by up to $1.35\times$ by enabling parallel query execution across DPU clusters, each holding a full copy of the DB, unlike the single cluster setup which processes queries serially on a DB split across all DPUs.}

\subsection{Comparison with GPU-based PIR}
\label{eval:gpu}
\begin{figure}[!t]
    \centering
    \includegraphics[width=1\linewidth]{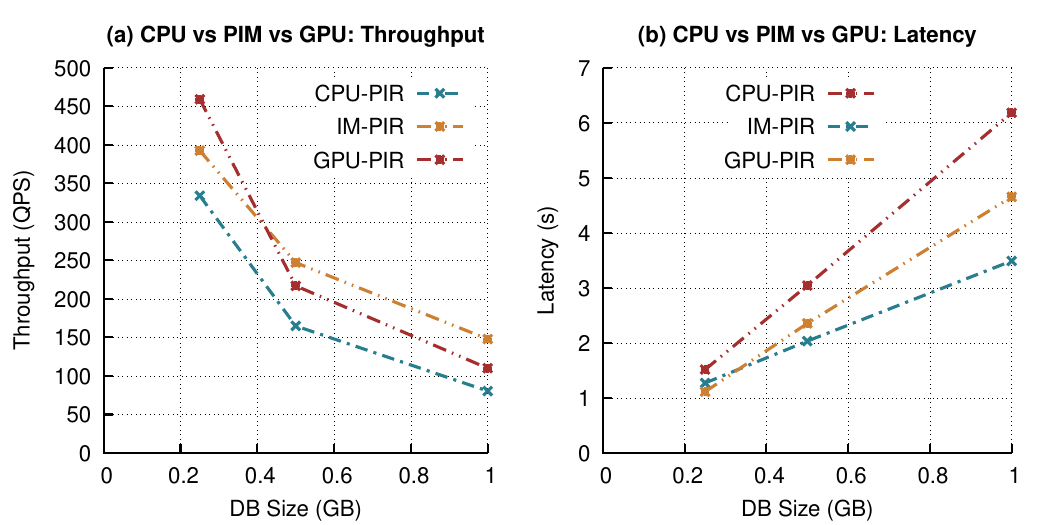}
    \caption{Throughput/latency comparison between \sys, CPU-PIR, and a GPU-based PIR from previous work \cite{GPU-DPF} on varying DB sizes. Overall, \sys shows both throughput and latency improvements with respect to these processor-centric approaches.}
    \label{fig:gpu}
\end{figure}

In this section, we aim to compare \sys to a GPU-based PIR approach. 
The latter is based on the PIR implementation from \cite{GPU-DPF} which equally uses a multi-server PIR construction based on distribute point functions.
The corresponding code for this work is open-source; we executed the code on a GPU for varing DB sizes.
\autoref{fig:gpu} illustrates the results obtained.

We observe that \mbox{\sys} achieves up to $1.34\times$ throughput and $1.3\times$ latency improvement compared to the GPU-based approach of \mbox{\cite{GPU-DPF}} (GPU-PIR).

Also, compared to the CPU-PIR, the GPU-based approach achieves up to $1.36\times$ throughput and $1.3\times$ latency improvement.
Though GPUs provide higher parallelism and memory bandwidth compared to CPUs, GPUs still follow a processor-centric approach and can be bottlenecked by limited memory bandwidth.
On the other hand, PIM-based setup provides even more parallelism, memory bandwidth, and better scaling of its compute resources with memory compared to the GPU-based setup.
Nevertheless, prior work \mbox{\cite{nider20}} shows that GPUs are expected to perform much better than DPUs on workloads where memory bandwidth is not the main bottleneck.

\observ{Compared to GPUs, PIM architectures like UPMEM's provide more parallelism, memory bandwidth, and better scaling of their compute resources with memory, allowing for higher query throughput improvements in multi-server PIR designs, especially when dealing with large databases.}

\section{Related work}
\label{sec:rw}

\subpoint{Single and multi-server PIR.}
Private information retrieval was first introduced by Chor et al. \mbox{\cite{chor95}}, which considered the multi-server model where the database is replicated across non-colluding servers.
Kushilevitz and Ostrovsky~\cite{single_pir} proposed the first single-server PIR construction based on additive homomorphic encryption, removing the reliance on non-colluding servers.
Building upon these works, several subsequent works proposed both single-server~\cite{simplePIR, OnionPIR, sealPIR, FastPIR, XPIR} and multi-server~\cite{fss, CA, RAID_PIR, popcorn} PIR schemes, providing algorithmic methods for improving both computational and communication complexity.
Specifically, several of these multi-server schemes have shown that the memory-bound XOR operations on the entire database are a major performance bottleneck.

\subpoint{Hardware-accelerated PIR.}
Several studies have proposed hardware-acceleration for both single- and multi-server PIR.

\subsubpoint{Single-server acceleration.}
Melchor \emph{et al.}~\cite{melchor08} proposed the first GPU-accelerated PIR scheme to mitigate the computational overhead in a lattice-based single-server PIR scheme~\cite{melchor2007lattice}. 
Maruseac \emph{et al.}~\cite{PIR_HE} leverage GPUs to speed up large integer multiplications and modulo products which form the basis of several single-server cryptographic algorithms.
Dai \emph{et al.}~\cite{dai2015accelerating} use GPUs to accelerate a somewhat homomorphic encryption (SWHE)-based single-server PIR scheme~\cite{doroz2014bandwidth}.
Their work offloads expensive modular multiplications and modulus switching, which are the main bottleneck of many single-server PIR protocols, to GPUs.
INSPIRE~\cite{inspire} proposes an in-storage (SSD) processing architecture to accelerate FHE in single-server PIR.
Our work shares similarities with this work as it addresses bandwidth limitations.
However, \sys leverages in-memory processing/PIM (not in-storage processing) in the multi-server setting to accelerate lightweight XOR operations; the PIM architecture we use is not adapted for FHE and single-server PIR in general.

\subsubpoint{Multi-server acceleration.}
G{\"u}nther \emph{et al.}~\cite{GPU-CIP} use GPUs to improve the computational costs of the XOR operations in a multi-server PIR setup.
They equally introduce client-independent preprocessing (CIP) which move some server computations to an offline phase.
Lam \emph{et al.}~\cite{GPU-DPF} propose a GPU-accelerated multi-server PIR scheme for machine learning inference.

Contrary to these works which leverage processor-centric compute architectures, our work introduces a memory-centric multi-server PIR design as a solution to memory-bound server-side operations (\eg XORs) which, as shown by previous studies, are a major bottleneck for multi-server PIR.
Our design mitigates the memory bandwidth limitations of processor-centric solutions based on GPUs and CPUs.

\subpoint{PIM-based acceleration for workloads.}
While our work is the first to propose PIM-based acceleration for multi-server PIR, several prior studies have showcased the advantages of PIM in accelerating data intensive workloads including graph processing~\cite{ahn-isca15, pimpam, huang20graphProcessing,tcim, asquini2025accelerating}, genomic analysis~\cite{alser2020accelerating, Hur2024AcceleratingDR, lavenier16}, machine learning~\cite{PIM-OPT,PyGIM,Reenforcement_Learning}, and database operations~\cite{pim-join, spid-join, Accelerating_large_table_scan}.
Recent studies~\cite{PIM_HE, gilbert24} explored PIM-based acceleration for fully homomorphic encryption and their results showed that current PIM architectures like UPMEM are not well adapted for FHE-based algorithms.

Our work leveraged the insights gained from these studies to propose a solution which mitigates the overhead of memory-bound multi-server PIR operations.

\section{Conclusion}
\label{sec:conclusion}
In this work, we show how the algorithmic foundations of (memory-bound) multi-server PIR computations align with the core strengths of processing-in-memory architectures, and propose \sys, the first multi-server PIR design based on real PIM hardware.
Our evaluation demonstrates that \sys significantly improves query throughput by more than $3.7\times$ when compared to a standard CPU-based PIR approach.
To the best of our knowledge, our work represents the first effort to accelerate PIR using PIM, and we believe our results mark an important step toward the broader adoption of PIM for PIR-based solutions.


	
	\bibliographystyle{ACM-Reference-Format}
	\bibliography{ref, pim_fhe, pim_cache}

\end{document}